\documentclass[conference,compsoc]{IEEEtran}

\ifCLASSOPTIONcompsoc
  \usepackage[nocompress]{cite}
\else
  \usepackage{cite}
\fi
\usepackage{amsmath,amssymb}
\usepackage{graphicx}
\usepackage{booktabs}
\usepackage{xcolor}
\usepackage{url}
\usepackage[hidelinks]{hyperref}
\hypersetup{pdftitle={Zero-Trust Formal Methods: AI Agents Build Verified Checkers for Untrusted Solvers},pdfauthor={Tobias Philipp}}
\usepackage{xspace}
\usepackage{tikz}
\usetikzlibrary{positioning,calc,arrows.meta}

\newif\ifdraft \draftfalse
\ifdraft
  \newcommand{\todo}[1]{{\color{red}\textbf{[TODO:} #1\textbf{]}}}
  \newcommand{\budget}[1]{{\color{blue}\footnotesize [Budget: #1]}}
\else
  \newcommand{\todo}[1]{}
  \newcommand{\budget}[1]{}
\fi

\makeatletter
\def\paragraph{\@startsection{paragraph}{4}{2\parindent}{0ex plus 0.1ex minus 0.1ex}{0ex}{\normalfont\normalsize\bfseries}}
\makeatother

\newcommand{\isabelle}{Isabelle/HOL\xspace}
\newcommand{\verus}{Verus\xspace}
\newcommand{\zthree}{z3\xspace}
\newcommand{\clingo}{clingo\xspace}
\newcommand{\clasp}{clasp\xspace}

\begin{document}
\bstctlcite{IEEEexample:BSTcontrol}%

\title{Zero-Trust Formal Methods:\\ AI Agents Build Verified Checkers for Untrusted Solvers}

\author{\IEEEauthorblockN{Tobias Philipp}
\IEEEauthorblockA{secunet Security Networks AG\\
Germany\\
tphilipp@gmail.com}}

\maketitle
\begin{tikzpicture}[remember picture, overlay]
  \node[anchor=south, text width=16cm, align=center, font=\footnotesize]
    at ([yshift=0.45in]current page.south)
    {This work has been submitted to the IEEE for possible publication. Copyright may be
     transferred without notice, after which this version may no longer be accessible.};
\end{tikzpicture}%

\begin{abstract}
AI coding agents now generate code faster than humans can review it. Security analyses rely on solver answers such as ``unsatisfiable'' (for example, no attack path exists) or ``optimal.'' Unlike satisfying models, they cannot be checked by looking at them. Formal verification moves trust from code to specifications, but verifiers still place large SAT, SMT, and answer-set solvers, and the translation into solver queries, in their trusted computing base, and these have had soundness bugs. Unlike agent-based pipelines that trust their verifier and solver, our farm trusts no component by default: neither the AI agents, which might make mistakes or game their objectives, nor the solvers. A solver answer counts only if a verified checker accepts its certificate. Every theory, proof, and line of code is generated by AI. For each component, agents first create an \isabelle theory with a specification, a soundness proof, and a reference checker made by code generation, then a fast Rust implementation, verified with \verus against a specification that mirrors the Isabelle definitions line by line. Specifications are frozen after independent review and cannot be changed by the implementing agents. A trust manifest lists what is still trusted. We built trust chains for SAT (verified LRAT checking); SMT, with certificates from an extended \zthree and verified bit-blasting for bit-vectors and floating point; answer-set programming, with native certificates from \clingo, including weight rules and optimality; and, closing the loop, \verus's final verification-condition step, by translation validation. All 3,726 SMT queries from 20 crates are certified and validated, and 15,662 answer-set proofs are accepted by both the Isabelle and the \verus checker. We found soundness defects in three proof checkers, including drat-trim, and in five solvers, including \clasp and \zthree, which report wrong unsatisfiability.
\end{abstract}

\section{Introduction}\label{sec:intro}

AI coding agents now produce code faster than humans can review it, and this code is often insecure~\cite{copilot,llmhardening,baxbench}. Formal verification offers a way out of this bottleneck: if a machine checks the code against a specification, humans need to review only the specification. Large-language-model-based (LLM-based) verification pipelines and benchmarks pair language models with program verifiers like \verus, Dafny, or Lean, and use the verifier's verdict, usually guarded by filters against verification bypasses, as the measure of success~\cite{verina,alphaverus,autoverus,clover,dafnybench}.

That verdict deserves less trust than it receives. A verifier like \verus does not check proofs itself. It translates a program and its specification into satisfiability modulo theories (SMT) queries and accepts the program if \zthree answers ``unsatisfiable'' to every query. The verdict therefore rests on the solver and on the translation, two large and fast-moving code bases; \zthree alone has a record of dozens of soundness bugs~\cite{yinyang}. An agent rewarded for passing the verifier has every reason to find such bugs, and to find weaker paths to a green check: a recent industrial report observed a coding agent that silenced failing proof obligations with an assumption and still reported success~\cite{proverjudge}, and an academic pipeline observed a model that learned to game the verifier with \texttt{assume(false)}~\cite{alphaverus}.

The same weakness appears wherever security analyses rest on automated solvers or provers. Many such analyses state their strongest results as negative answers: no constant-time violation in a binary, up to a bound~\cite{binsecrel}, no speculative leak within a bounded speculation window~\cite{spectector}, no timing leak in a hardware design~\cite{xenon}, no attack on a protocol in the symbolic model~\cite{proverif}. When a solver reports that a formula is satisfiable, it can return a model that anyone can check. When it reports ``unsatisfiable'' (no attack path exists, no input violates the policy) or ``optimal'' (no cheaper configuration exists), there is nothing to inspect. A wrong answer of this kind is a spurious proof that the analysis accepts, and it produces false assurance exactly where assurance matters.

Certificates address this weakness. A solver can emit a proof of unsatisfiability, and a small checker can validate it. For Boolean satisfiability (SAT), fast, formally verified checkers exist for hinted clausal proofs in linear resolution asymmetric tautology (LRAT) format or the similar GRAT format~\cite{lrat,grat,cakelpr}. For SMT, proof formats and checkers exist, but the checkers are unverified and skip some steps~\cite{carcara}, or cover only part of the solver's rules, theories, or preprocessing~\cite{leansmt,dratsmt}. For answer-set programming (ASP), inconsistency proofs exist~\cite{aspdrupe,aspqrat}, but ASP-DRUPE handles weight rules only by expanding them into normal rules and treats optimality only as a sketch; ASP-QRAT covers neither, and no production solver emits them in its default configuration. Agentic pipelines use none of these. They trust the verifier and its solver.

\paragraph*{Zero trust for formal methods}
In network security, \emph{zero trust} means that no component is trusted because of where it sits: trust is never granted implicitly but must be continually evaluated, and every access request is authenticated and authorized~\cite{nist80020}. We apply this principle to formal verification. No component of the verification pipeline is trusted implicitly: not the AI agents, which may make mistakes or try to game their objectives; not the solvers; and not the translation into solver queries. What remains trusted is listed in a manifest.

\paragraph*{Our approach}
Our target is security-critical software, such as cryptographic libraries, parsers, and protocol implementations, whose assurance ultimately rests on formal methods, that is, on solver answers and verifier verdicts that cannot be inspected. That is precisely why we make the formal-methods layer itself trustworthy rather than trusting it. We built a \emph{verification farm} in which AI agents write every theory, proof, and line of code. A human states the intent and reviews the specifications for adequacy. No human reviews the generated code or proofs. A single proof-assistant kernel, \isabelle, anchors all specifications. For each checker, the agents first develop an Isabelle theory with a specification, a soundness proof, and a reference checker obtained by code generation. They then build a fast Rust checker, verified with \verus against a specification that mirrors the Isabelle definitions line by line. Specifications are frozen after review and protected from the agents that implement them. We accept a solver answer only if a verified checker approves its certificate.

The checkers are themselves products of the farm, and their own verification rests on \zthree through \verus. We close this loop: every \zthree answer that \verus obtains while verifying the three checker crates is certified by the farm's checkers (the proofs inside \verus's standard library remain trusted), and the last step of \verus's translation into SMT is validated against an Isabelle semantics.

\paragraph*{Contributions}
\begin{itemize}
  \item A zero-trust architecture for agentic verification: one \isabelle anchor, two separately implemented verified checkers per certificate format (except the bit-vector and floating-point chain, which has one), specification integrity enforced by the environment, and trust manifests (Sect.~\ref{sec:farm}).
  \item A certificate format for \zthree that combines LRAT-style hints with checkable theory steps, two verified checkers for it, and a verified bit-blasting chain for bit-vectors and IEEE floating point (Sect.~\ref{sec:satsmt}).
  \item Native certificates from \clingo in its default configuration, covering inconsistency, weight rules, and optimality, with two verified checkers (Sect.~\ref{sec:asp}).
  \item Certification of every solver answer \verus relies on, and translation validation of its final verification-condition step (Sect.~\ref{sec:verus}).
  \item An evaluation: all 3,726 SMT queries from 20 crates (our own example crates and the three checkers) are certified and validated, both ASP checkers accept 15,662 answer-set proofs, and the verified LRAT checker matches the fastest unverified checker on one thread and is faster on eight (Sect.~\ref{sec:eval}).
  \item A red-team evaluation of the farm's own gates with seeded defects: the verified checkers stopped every attack, while the tools that enforce the process leaked in six places, which we fixed and re-measured (Sect.~\ref{sec:redteam}).
  \item Soundness defects in three proof checkers, including drat-trim; in \clasp, \zthree, and two bit-vector solvers, which report wrong unsatisfiability; and in a proof-producing SMT solver whose valid proofs refute a problem other than its input (Sect.~\ref{sec:defects}).
\end{itemize}
\section{Background}\label{sec:background}

\subsection{Certificates for unsatisfiability}
A propositional formula in conjunctive normal form (CNF) is a set of clauses. A \emph{clausal proof} of unsatisfiability lists clauses, each of which follows from the formula and the earlier clauses by a simple check, and ends with the empty clause. The simplest check is \emph{reverse unit propagation} (RUP): assume the negation of the new clause, propagate unit clauses, and reach a conflict. In deletion resolution asymmetric tautology (DRAT) proofs, the checker must search for the clauses that propagate. LRAT proofs~\cite{lrat} name them: each step carries a \emph{hint}, the list of earlier clause identifiers that become unit, in order, the last one falsified. For example, the clauses (1)~$x\lor y$, (2)~$x\lor\neg y$, (3)~$\neg x\lor y$, and (4)~$\neg x\lor\neg y$ are refuted by two steps. Step~5 adds the clause $x$ with hint $1,2$: assuming $\neg x$, clause~1 forces $y$, and clause~2 is then false. Step~6 adds the empty clause with hint $5,3,4$: clause~5 forces $x$, clause~3 forces $y$, and clause~4 is false. A checker for such hinted proofs (LRAT, or the similar GRAT format) therefore performs no search, which makes it small enough to verify~\cite{lrat,grat,cakelpr}. LRAT also allows \emph{resolution asymmetric tautology} (RAT) steps, which preserve satisfiability rather than logical implication. Deletion steps remove clauses that are no longer needed.

\subsection{SMT and theory reasoning}
SMT solvers decide formulas over theories such as linear arithmetic, uninterpreted functions, and datatypes. In the architecture of conflict-driven clause learning (CDCL) modulo theories, CDCL(T), a SAT solver works on the Boolean abstraction of the formula. Theory solvers add \emph{theory lemmas}: clauses that are valid in the theory, such as $\neg(x>0)\lor\neg(x<0)$. Before solving, the solver preprocesses the input by rewriting terms, introducing Skolem functions, and instantiating quantifiers. A certificate for an SMT answer must justify all of this, not only the propositional reasoning. Existing approaches include the Alethe format with an elaborating checker~\cite{carcara}, reconstruction in interactive provers~\cite{leansmt}, and DRAT-style proofs extended with theory lemmas~\cite{dratsmt}.

\subsection{Answer-set programming}
An answer-set program is a set of rules $h \leftarrow b_1,\dots,b_n$, possibly with negation, choice, and weight constraints. Its answer sets are the models in which every true atom is derived without circular support. Checking a claimed answer set is easy. Showing that no answer set exists, or that an answer set is optimal under a minimize statement, is not: a solver such as \clingo~\cite{clingo} translates rules into constraints, learns clauses, and refutes \emph{unfounded sets}, sets of atoms that could only support each other~\cite{cdnl}. The completion of a program together with loop formulas characterizes its answer sets~\cite{assat,loopformulas}. ASP has been used to find attacks on security protocols~\cite{aspprotocols}. In such analyses, ``no attack exists'' (up to a bound) is a claim that no answer set exists~\cite{aspprotocols}, and ``this set of countermeasures is best'' is a claim of optimality.

\subsection{\verus and \isabelle}
\verus~\cite{verus} verifies Rust programs annotated with specifications and proofs. It translates a program, via the Rust compiler's high-level intermediate representation (HIR), into its intermediate language VIR, then into AIR, an SMT-oriented language with statements and assertions, and finally into SMT-LIB queries for \zthree. \verus accepts a function if \zthree answers ``unsat'' for each of its queries~\cite{verus}. Proofs marked \texttt{integer\_ring} go to Singular instead, and our crates use none. \isabelle is a proof assistant for higher-order logic (HOL) in the style of the Logic for Computable Functions (LCF) system: every theorem is checked by a small kernel. Its code generator exports executable functions as Standard ML (SML), which we compile with MLton.
\section{Threat Model and Goals}\label{sec:threat}

\paragraph*{Untrusted agents}
AI agents write all code, proofs, and theories, and we assume that any of them may be wrong or may game its objective. A gaming agent can weaken a specification until it becomes trivial; add verification bypasses such as \texttt{sorry} or an axiom in Isabelle, or \texttt{assume}, \texttt{admit}, or \texttt{external\_body} in \verus; edit tests or gates; exploit a bug in a checker; or report success that no check supports. We do not assume that agents are malicious in a targeted sense, but the design must hold if an agent behaves like an adversary that is rewarded for a passing verdict.

\paragraph*{Untrusted solvers and translations}
SAT, SMT, and ASP solvers may return wrong answers: a wrong ``unsat'', a wrong ``optimal'', or a non-model printed as a model. The translation from a program to solver queries may also be wrong. A certificate may be malformed or crafted. None of these components may be able to make the farm accept a false claim.

\paragraph*{What remains trusted}
We trust the \isabelle kernel, its code generator, and MLton for the reference checkers; \verus's front end (Rust to VIR to AIR) and \texttt{rustc} for the fast checkers; a small set of parsers that read input files; the I/O wrappers that print verdicts; the human review of specifications; and, for the artifacts the farm produces, the tools that enforce specification integrity and check the manifests. These tools run under the agents' own operating-system account: they stop the attacks we tested, but they do not exclude an agent that deliberately disables them. Appendix~\ref{app:manifest} lists the trusted base in full and gives its size. We do not consider attacks on hardware, the operating system, or the build machine.

\paragraph*{Goal}
If the farm reports a property as \emph{certified}, the property holds for the input as read by the trusted parser, relative to the frozen specification, assuming only the trust manifest. A \emph{not certified} verdict carries no claim.

\paragraph*{Non-goals}
Freezing a specification protects its \emph{integrity}: after review, the environment refuses an implementing agent's changes to it, within limits that we measure. It does not guarantee its \emph{correctness}. A wrong specification can be frozen. We address correctness by review and by testing the specification against the solvers' observable behavior, but we cannot prove it. Completeness is also not a goal: a checker may reject a correct certificate, which costs coverage, never soundness.
\section{The Verification Farm}\label{sec:farm}

\begin{figure}[t]
\centering
\begin{tikzpicture}[
  font=\scriptsize,
  box/.style={draw, rounded corners=2pt, align=center, inner xsep=3pt, inner ysep=2pt, minimum height=8mm},
  trusted/.style={box, fill=black!10},
  untrusted/.style={box, dashed},
  arr/.style={-{Stealth[length=1.6mm]}, semithick},
  junction/.style={circle, fill, inner sep=0.8pt}]
  \node[trusted, minimum width=17mm] (spec) at (0.88,0) {Isabelle/HOL\\specification\\(reviewed, frozen)};
  \node[untrusted, minimum width=18.5mm] (thy) at (3.235,0) {soundness proof\\(agents)};
  \node[box, minimum width=20mm] (ref) at (5.61,0) {reference checker\\(code generation)};
  \node[box, minimum width=17mm] (human) at (0.88,-1.35) {human:\\intent, review};
  \node[untrusted, minimum width=18.5mm] (solver) at (3.235,-1.35) {\zthree, \clingo,\\CaDiCaL, \verus};
  \node[untrusted, minimum width=20mm] (cert) at (5.61,-1.35) {certificate};
  \node[box, minimum width=14mm] (verdict) at (7.69,-1.35) {certified /\\not certified};
  \node[untrusted, minimum width=18.5mm] (mirror) at (3.235,-2.7) {\verus mirror\\(agents)};
  \node[box, minimum width=20mm] (fast) at (5.61,-2.7) {fast checker\\(Rust, verified)};
  \draw[arr] (human) -- (spec);
  \coordinate (split) at ($(spec.east)!0.45!(thy.west)$);
  \draw[arr] (spec.east) -- (thy.west);
  \draw[arr] (split) |- (mirror.west);
  \node[junction] at (split) {};
  \draw[arr] (thy) -- (ref);
  \draw[arr] (mirror) -- (fast);
  \draw[arr] (solver) -- (cert);
  \draw[arr] (cert) -- (ref);
  \draw[arr] (cert) -- (fast);
  \draw[arr] (ref.east) -| (verdict.north);
  \draw[arr] (fast.east) -| (verdict.south);
\end{tikzpicture}
\caption{Overview. Dashed components are untrusted: agents, solvers, and certificates. A specification enters the trusted base only through human review. Each certificate format has two checkers, both proved sound against the same specification. The fast checker's proof refers to its line-by-line \verus mirror.}
\label{fig:overview}
\end{figure}
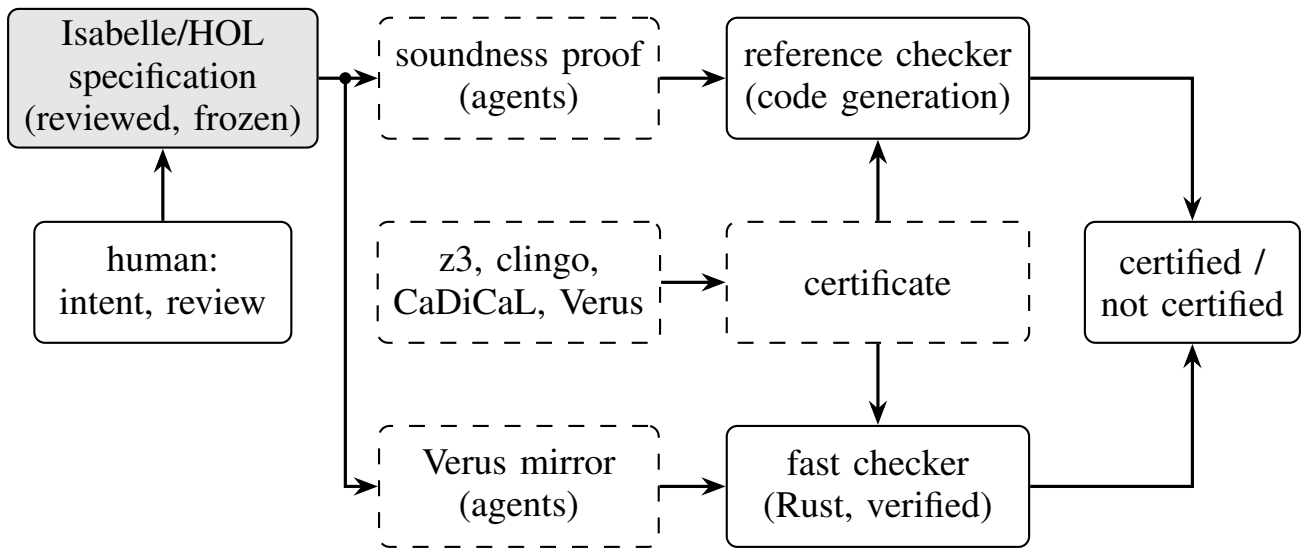

Figure~\ref{fig:overview} gives an overview. The farm is a set of principles, enforced by tools, under which AI agents produce verified software.

\subsection{Human intent, machine checking}
The farm is AI-first. A human decides what to build and reviews each specification for adequacy: does it say what is meant? No human reviews generated code or proofs. We accept a proof because the Isabelle kernel or \verus accepts it. We accept a solver answer because a verified checker accepts its certificate. Code that is trusted but not verified, such as the parsers that read input files, is not reviewed by a human either. Instead, it is covered by independent implementations and by differential and mutation testing, and it is listed in the trust manifest.

\subsection{One anchor, two checkers}
Every specification lives in \isabelle first. Every other tool either produces certificates that are checked against an Isabelle specification or is listed as trusted. For each certificate format, the agents first write an Isabelle theory: the semantics, the checking rules, and a proof that a checker that accepts a certificate implies the claim, for example that the input formula is unsatisfiable (Appendix~\ref{app:theorems} states each top-level theorem). Isabelle's code generator turns the checker into a \emph{reference checker}.

The agents then write a second, fast checker in Rust and verify it with \verus. Its trusted specification mirrors the Isabelle definitions line by line, and its proof mirrors the structure of the Isabelle soundness proof. Separate agents write the two checkers in separate working copies, with separate data structures and proofs, checked by different tools that rest on different solvers and kernels. The two are not isolated: the \verus agents read the Isabelle theories they mirror, and the \verus ASP parser follows the Standard ML parser. For \verus queries, the driver counts a query as certified only if both SMT checkers accept it or the bit-blasting chain does. For SAT and ASP, each checker is a standalone tool with its own theorem, and a false certificate must fool both whenever both are run. Strictly, the fast checker is proved against the \verus mirror. That the mirror corresponds to the Isabelle definitions is reviewed and tested differentially, not proved. Differential testing between the two checkers validates the mirror and finds completeness gaps.

\subsection{Specification integrity}
A specification is frozen after independent review: by the human for intent, and by reviewing agents that search for counterexamples, which become must-reject tests. A frozen specification is protected from modification by the agents that implement against it. The protection is enforced by the environment, not by instruction: a hook refuses an implementing agent's writes to frozen files through editing tools and common shell commands. An implementing agent that finds a specification unworkable may report a conflict with evidence but cannot resolve it by editing the specification.

\subsection{Trust manifests}
Every artifact carries a manifest listing what remains trusted: each verification bypass in a proof (\texttt{sorry}, axioms, \texttt{assume}, \texttt{admit}, \texttt{external\_body}), each trusted parser, each I/O wrapper, and each tool whose output is not checked, each with a justification. A hygiene tool compares the verification bypasses in the code with the manifest and fails on any difference. The farm never reports ``verified'' without its manifest. Manifests play the role that attestations play in supply-chain security~\cite{intoto}: they state, per artifact, what a consumer must still trust. Unlike those attestations, our manifests are not yet signed and bound to a binary digest by a party outside the agents' environment.

\begin{table}[t]
\centering
\caption{Threats and the mechanisms that address them.}
\label{tab:threats}
\footnotesize
\begin{tabular*}{\columnwidth}{@{}l@{\extracolsep{\fill}}l@{}}
\toprule
Threat & Mechanism \\
\midrule
Agent weakens a specification & Freeze after review, refuse edits \\
Agent hides a verification bypass & Hygiene check against the manifest \\
Agent edits gates, fakes success & Gates re-run by a separate agent \\
Agent exploits a checker bug & Two verified checkers per format \\
Checker prints a wrong verdict & Verified boundary, frozen I/O \\
Solver errs on unsat or optimum & Certified by verified checkers \\
Certificate is crafted or broken & Untrusted, bad steps only fail \\
Verifier mistranslates code & AIR$\to$SMT validated per query \\
Two trusted readers disagree & Differential tests on tokens, trees \\
\bottomrule
\end{tabular*}
\end{table}

\subsection{Independent gates}
Evidence produced by an agent about its own work does not count. Before integrating work, a separate agent re-runs every gate from a clean state: clean rebuilds, reproducible code exports, bit-for-bit comparison of binaries with the producer's builds~\cite{reprobuilds}, test suites, and often an additional counterexample written for the purpose. Default tools and binaries change only after a full recount.
\section{Certified SAT and SMT}\label{sec:satsmt}

\subsection{LRAT checking}
The Isabelle theory defines CNF semantics and LRAT steps and proves that an accepted proof implies unsatisfiability. Its reference checker is \texttt{lrat-isa-ref}. The \verus checker proves the end-to-end statement
\[
\mathit{accepted} \Longrightarrow \exists f.\ \mathit{parse}(b) = \mathit{Some}\ f \land \neg\mathit{sat}(f),
\]
where $b$ is the input byte string and $\mathit{parse}$ is a verified DIMACS parser checked against a reviewed byte-level grammar. Three design choices improve speed. First, a verified bijection renames sparse variable numbers, so all tables are bounded by the input size. Second, for proofs that use only RUP, the checker verifies lemmas independently and in parallel. A lemma proved in Isabelle and mirrored in \verus states that the conjunction of these local checks implies unsatisfiability. Third, on any anomaly, such as a RAT step, a hint that re-adds a still-live clause identifier, a hint that names an unknown identifier, or malformed data, the checker falls back to an exact sequential backward check.

The verdict boundary is also verified. The function that prints ``verified'' and exits requires the soundness claim over an uninterpreted name for the input bytes, and every exit function ensures \texttt{false}. A program that swaps or deletes an exit call no longer verifies. The bodies of the exit functions, which print the verdict line and set the exit code, remain trusted I/O. The manifest freezes them by hash, together with all other trusted I/O code. Mutation testing showed why this matters: before the change, a version of the checker with swapped exit calls verified and printed ``verified'' on every rejection.

\subsection{SMT certificates from \zthree}
We extended \zthree to write a certificate with each ``unsat'' answer. A certificate is a sequence of clausal steps. Each step adds a clause with an identifier and a justification. Deletion steps remove clauses. A step is justified in one of three ways:
\begin{itemize}
  \item \emph{Propositional:} \texttt{Rup} with an LRAT-style hint, the clauses that become unit in order; \texttt{Tseitin} for the clauses that define a Boolean gate (16 templates).
  \item \emph{Theory:} \texttt{Cong} with an explicit equality derivation (reflexivity, symmetry, transitivity, congruence); \texttt{Farkas} with one rational coefficient per literal for linear arithmetic; \texttt{Tri} for implied arithmetic equalities; \texttt{Bound} for integer rounding; \texttt{Inst} for quantifier instances; and \texttt{Axiom} for instances of named schemas, including datatype axioms, integer division, and a sign rule for products that covers the nonlinear lemmas \zthree generates.
  \item \emph{Definitional:} \texttt{Rewrite}, checked by a verified normalizer that brings both sides into polynomial normal form, and \texttt{Define} and \texttt{Skolem}, which extend the signature with symbols that occur nowhere else.
\end{itemize}
Steps that \zthree cannot yet justify are written as explicit \emph{holes}. A hole always fails its check. Figure~\ref{fig:smtcert} shows a complete certificate. Appendix~\ref{app:formats} gives the full format.

\begin{figure}[t]
\footnotesize
\begin{verbatim}
(assume $24)                  ; 1 a = b
(assume $33)                  ; 2 p(a)
(assume $45)                  ; 3 not p(b)
(infer (not $45) (not $44)
  (tseitin $45 not-pos))      ; 4 not-gate
(infer (not $44) (rup 3 4))   ; 5 not p(b)
(infer $47 (not $24)
  (cong $47 (prem $24)
    (cc $33 $44 0)))          ; 6 congruence
(infer (not $47) (not $33) $44
  (tseitin $47 iff 0))        ; 7 iff gate
(infer (rup 1 2 6 7 5))       ; 8 empty clause
\end{verbatim}
\caption{Certificate for the query $a=b,\ p(a),\ \neg p(b)$. \texttt{\$k} refers to a node of the term table shared with the input. All three assumptions are live input assertions, and the cone is steps 1--8, so the verdict is \emph{certified}.}
\label{fig:smtcert}
\end{figure}

The checker computes the \emph{cone} of each query: the steps reachable from the empty clause through hints. Only cone steps must pass. Steps outside the cone may fail without effect. The verdict is \emph{certified} if every cone step passes and every assumption in the cone is a live input assertion of the query; \emph{certified-relative} if some assumption is not an input assertion, in which case only the listed assumptions are refuted, and this verdict never counts as certified; and \emph{not certified} otherwise. The live-assertion rule binds a certificate to the input as the trusted parser reads it, including its \texttt{push} and \texttt{pop} scopes.

The semantics is many-sorted first-order logic with Booleans, integers, reals, and uninterpreted sorts with arbitrary nonempty carriers. The \verus soundness theorem is generic in the carrier type and in any ordered field for the reals. The trusted parsers accept only a closed subset of SMT-LIB: the subset that \verus emits. RAT steps are known to be unsound modulo theories, so the format excludes them. Fresh symbols enter only through \texttt{Define} and \texttt{Skolem} steps, and the checker verifies their shape.

\subsection{Two checkers}
Both checkers implement the same rules against the same Isabelle semantics, but with separate parsers and separate rule implementations: the reference checker \texttt{smt-isa-ref} is exported from Isabelle, and \texttt{smt-check} is written in \verus. Its verification comprises 1,394 SMT queries over 1,145 functions. Every \verus rule check is proved correct against its mirror of the Isabelle rule, and the mirrors are proved sound. An Isabelle theorem connects the formats: an accepted SMT \texttt{Rup} step is an accepted LRAT step under the obvious encoding of literals. The two trusted parsers are cross-checked by comparing canonical dumps of their parse trees.

\subsection{Bit-vectors and floating point (BV/FP)}
Both SMT checkers reject bit-vector queries. These go to a separate chain in Isabelle that bit-blasts the query to CNF and has CaDiCaL produce an LRAT proof, which the verified LRAT checker inside the same chain checks. A single Isabelle theorem covers the path from the SMT-LIB input to the LRAT check: the CNF is unsatisfiable exactly when the bit-vector formula is. The chain covers arithmetic, including multiplication, division, and remainder; shifts; signed operators; extraction; concatenation; and extension.

Floating point follows the SMT-LIB theory, which is based on the formal semantics of Brain et al.~\cite{fpsmt}. Our chain has circuits for unpacking, classification, comparison, rounding, addition, subtraction, and multiplication, proved correct against an IEEE-754 reference model under round-to-nearest-even. Other rounding modes yield \emph{not certified}. One detail illustrates the subtlety: not-a-number (NaN) has several bit patterns~\cite{fpsmt}, and under \zthree's reading of \texttt{fp.to\_ieee\_bv} a fixed bit pattern for NaN is unsound, so the blaster introduces a fresh, NaN-constrained vector per symbol. CaDiCaL is never trusted.
\section{Certified Answer-Set Programming}\label{sec:asp}

ASP solvers report three kinds of answers that matter here: an answer set, ``no answer set'' (inconsistency), and ``this answer set is optimal''. Only the first can be checked by inspection. We designed a native certificate format that \clingo emits while it solves, rather than translating the program into another formalism.

\subsection{Certificate format}
The checker does not trust the solver's internal representation. It reads the ground program in \clingo's intermediate format and computes the \emph{clause view} itself: the Clark completion over atom and body variables. A certificate uses these steps:
\begin{itemize}
  \item \texttt{a}: a clause added by RUP with an LRAT-style hint; \texttt{d}: deletion.
  \item \texttt{u}: an \emph{unfounded-set step}. For any nonempty set $U$ of atoms, the step adds a clause that contains the loop formula of $U$. Its validity follows from the loop-formula theorem~\cite{loopformulas}, proved in Isabelle.
  \item \texttt{w}: a clause implied by the definition of one weight or cardinality body, with cover sets for unfounded-set steps over weight rules.
  \item \texttt{e}: a purely definitional extension, which introduces a fresh variable.
  \item \texttt{m}: a claimed answer set. A non-final \texttt{m} sets a bound for optimization, and \texttt{o} adds a clause implied by ``the cost is lexicographically below the bound''. The final verdict is \emph{optimum}.
  \item Cutting-plane steps over pseudo-Boolean constraints, in the style of VeriPB~\cite{veripbparity}, for the counters that \clingo introduces when it rewrites weight rules, without redundancy rules.
\end{itemize}
The verdicts are \emph{inconsistent} (no answer set exists), \emph{answer set}, and \emph{optimum} (the reported answer set is optimal). Our format builds on ASP-DRUPE~\cite{aspdrupe}, which introduced a proof format for checking ASP solvers' inconsistency claims, with an extension step. ASP-DRUPE treats weight rules as shorthand for normal rules and sketches optimality proofs as an answer set of cost $c$ together with an inconsistency proof for cost $c-1$. We give weight rules native steps and implement optimality proofs in the solver. ASP-QRAT~\cite{aspqrat} takes a different route through quantified Boolean formulas. It omits weight rules and is not implemented. As ASP-DRUPE observed, RAT steps combined with deletion and loop nogoods are unsound for non-tight programs, so our format introduces new variables only by definitions, whose shape the checker verifies.

\subsection{Proofs from \clingo in its default configuration}
The agents instrumented \clingo~5.8.3 (\clasp~3.4.2) in seven increments. The first emitted proofs for a restricted configuration. Later increments restored all preprocessing, added weight rules, definitional extensions for \clasp's internal rewrites, optimization, and cutting planes. The final version emits proofs in \clasp's default configuration. Each increment passed a gate in which both verified checkers accepted every proof of a large benchmark set, and in which the solver's answers and statistics were identical to those of unmodified \clingo. Proof output costs about $1.16\times$ the solving time.

Some configurations are still refused, for example recursive weight rules whose counters would need an unfounded-set step with a pseudo-Boolean cover. Refusal is safe: the checker reports no verdict.

\subsection{Two checkers}
The Isabelle theory comes in layers, one per extension, and each layer proves the three verdict claims. Each layer also proves an embedding theorem: certificates in the older format keep their verdicts. The semantics was validated against \clingo: Isabelle's evaluator computed the answer sets of 615 programs with at most four atoms, written by a reviewing agent, and of 343 programs with weight bodies and minimize statements (34 edge cases, 300 random, 9 with minimize), and agreed with the solver on every one. The \verus checker \texttt{asp-check} mirrors the Isabelle specifications line by line, including the trusted parser specification, and proves the same verdict claims over the parse of the input bytes.
\section{Closing the Loop: Certifying \verus}\label{sec:verus}

\verus verifies the fast checkers, and \verus trusts \zthree. Without further work, the farm would rest on the solver it set out to distrust. We remove this dependency in two steps: we certify every solver answer, and we validate the last step of the translation that produces the solver queries.

\subsection{Certifying every solver answer}
A driver runs \verus on a crate with the extended \zthree, collects every SMT query and the certificate for each ``unsat'' answer, and runs both SMT checkers on each query. Bit-vector queries go to the bit-blasting chain. The pairing of queries with solver processes is untrusted: a wrong pairing can only make a check fail. The result is a per-query verdict and a per-function summary. A query counts as certified only if the checkers accept, never because \zthree said so.

\subsection{Validating the final verification-condition step}
Certifying a query proves that the query is unsatisfiable. It does not prove that the query means what \verus intended. \verus translates AIR into SMT-LIB in two phases: AIR statements become a verification condition through static single assignment and a labeled weakest precondition, and the resulting AIR is printed as SMT-LIB. We validate these two phases per query.

The agents wrote a big-step semantics of AIR statements in Isabelle and proved that a valid verification condition implies that the AIR query is correct. A validator, exported from Isabelle, recomputes the verification condition from \verus's AIR log and compares it with the SMT query \verus actually sent to \zthree, up to renaming of versions, labels, and markers. The validator then emits the exact text it validated, and the certification step certifies exactly that text. The reported verdict is \emph{certified and validated on the same text}.

\subsection{Independent review finds a spurious proof}
The AIR semantics went through six independent reviews before it was frozen. Four failed. The validator reported full success in every round. The reviewers found, with counterexample lemmas or end-to-end files, spurious axioms that the semantics admitted for lifted lambda terms, atoms captured by enclosing binders, an alias between quoted and unquoted symbol names, and, in the fourth round, an end-to-end exploit: the trusted reader tokenized a crafted symbol differently from the SMT checker's parser, so that a spurious query was reported as both validated and certified. The fix makes the reader use the parser's token classes, and a standing differential test now compares token streams and parse trees on all corpus files plus one million fuzzed inputs and one million mutants, with no disagreements. The exploit is kept as an end-to-end must-reject test. This episode is the strongest evidence we have for the design: the agents' own tests passed, and only an independent reviewer with a specific counterexample found the hole.

\subsection{What remains}
\verus's front end, from Rust through VIR to AIR, remains trusted. Mutation testing gives indirect evidence: of 500 mutants of verified code, 492 compiled and 396 of them (80.5\%) were killed, and no surviving mutant pointed to a bug in verification-condition generation. Mutating verified code does not exercise the front end's translation directly. Closing this gap requires a semantics of the source language.
\section{Evaluation}\label{sec:eval}

We ask four questions. \textbf{RQ1}: What fraction of real solver answers can the farm certify? \textbf{RQ2}: What does certification cost? \textbf{RQ3}: How robust are the checkers and the trusted parsers that no human reviewed? \textbf{RQ4}: What did the agents produce, and what did the gates catch?

\paragraph*{Setup}
All experiments ran on one laptop: a MacBook with an Apple M5 Pro (15 cores) and 24\,GB of memory, running macOS~27.2. We used \isabelle 2025-2 with MLton 20241230, \verus 0.2026.10.30 (commit 3102966), \zthree 5.3.4 (development version, commit \texttt{deef01f}) with our certificate output, \clingo 5.8.3 with \clasp 3.4.2 and our proof output, and CaDiCaL 3.0.1. Reference checkers were lrat-trim 0.2.0, drat-trim's \texttt{lrat-check}, and cake\_lpr.

\begin{table}[t]
\centering
\caption{Coverage (RQ1). ``Both'': accepted by both checkers; ``BV'': bit-vector queries certified by the bit-blasting chain.}
\label{tab:coverage}
\footnotesize
\setlength{\tabcolsep}{4pt}
\begin{tabular}{@{}lrl@{}}
\toprule
Corpus & Certified & Notes \\
\midrule
\verus queries, 20 crates & 3,726/3,726 & 3,710 both, 16 BV \\
\quad functions & 3,282/3,282 & incl.\ three checkers \\
\quad validated (AIR$\to$SMT) & 3,726/3,726 & same text as certified \\
Bit-vector corpus & 80/81 & remaining one is satisfiable \\
LRAT corpus (21\,GB) & 206/206 & proofs up to 4.6\,GB \\
\clingo proofs & 15,662/15,693 & both; 100 option sets \\
\bottomrule
\end{tabular}
\end{table}

\subsection{RQ1: Coverage}
Table~\ref{tab:coverage} summarizes coverage. The \verus corpus consists of 17 example crates that cover contracts, loops, recursion, quantifiers, arithmetic, collections, traits, and refinement, plus the crates of the three \verus checkers themselves. The example crates are not third-party code: nine come from an earlier \verus tutorial of ours, and eight are probes that an agent wrote to survey \verus's features before the SMT format was designed. Coverage on this corpus shows that the format covers the code it was built for. It does not estimate coverage for arbitrary \verus code. For all 3,726 queries, both certification and translation validation succeeded on the same text. Both SMT checkers certified 3,710 queries, and the bit-blasting chain certified 16 bit-vector queries. \zthree judges no query. All four groups ran at the same checker sources with the same validator binary. The answer-set proofs come from 16,015 runs of \clingo with 100 option sets, the default configuration among them, on about 3,400 ground programs from the benchmark families that served as the acceptance gate during development. 322 runs refused proof output for options the format does not support. Of the remaining 15,693 runs (5,874 inconsistent, 2,987 optimum, 6,832 answer set), both ASP checkers accepted 15,662 proofs; 31 runs produced no valid proof. In the default configuration alone, 3,521 of 3,530 runs yielded accepted proofs. The 16 bit-vector queries counted above arise within the 20 crates. A separate bit-vector corpus of 81 queries, logged by \verus from survey probes, the example crates, and the checkers, exercises the bit-blasting chain directly, with 80 certified and the one remaining genuinely satisfiable. The 206 LRAT proofs come from CaDiCaL~3.0.1 (text and binary LRAT) and from kissat DRAT proofs converted by drat-trim, for 40 crafted instances and 50 instances drawn from the Global Benchmark Database with a fixed seed, four per family. The coverage figures do not benchmark SMT-LIB in general: the trusted SMT parsers accept the subset \verus emits, and queries outside it are reported as not certified.

\begin{table}[t]
\centering
\caption{\verus queries per crate: certified and validated, by both SMT checkers or by the bit-blasting chain (BV), and wall-clock time for \verus versus certification plus validation (4 threads). Times are upper bounds, measured with other work on the machine.}
\label{tab:crates}
\footnotesize
\setlength{\tabcolsep}{4pt}
\begin{tabular}{@{}lrrrrr@{}}
\toprule
Crate & Queries & Fns & Both/BV & \verus & Check \\
\midrule
ASP checker & 1,944 & 1,817 & 1,940/4 & 75.0\,s & 544.9\,s \\
SMT checker & 1,394 & 1,145 & 1,387/7 & 1,290.9\,s & 743.2\,s \\
LRAT checker & 195 & 153 & 193/2 & 17.7\,s & 31.6\,s \\
17 example crates & 193 & 167 & 190/3 & 21.1\,s & 25.6\,s \\
\midrule
Total & 3,726 & 3,282 & 3,710/16 & 1,404.7\,s & 1,345.3\,s \\
\bottomrule
\end{tabular}
\end{table}

\subsection{RQ2: Cost}
Verification need not be slow. On the 65 proofs of at least 10\,MB in the LRAT corpus, from 27 Global Benchmark Database instances in 14 families and 8 crafted instances, up to 4.6\,GB, the verified \verus LRAT checker matches the unverified lrat-trim single-threaded and needs 0.38 of its time with eight threads (64 proofs, as lrat-trim rejects the one with RAT steps). It is about twelve times faster than drat-trim's \texttt{lrat-check} and about eight times faster than the verified checker cake\_lpr, which we ran on the 29 proofs up to 150\,MB. cake\_lpr's guarantee extends to its machine code through CakeML's verified compiler, whereas our checker trusts \texttt{rustc}. The comparison is one of speed at different trust levels. Peak memory is comparable to lrat-trim's. All timings come from one laptop under concurrent load and are indicative rather than controlled measurements. The LRAT ratios are per-instance medians of three interleaved runs.

Certifying and validating all 3,726 \verus queries takes about as long as \verus's own verification of the same crates (Table~\ref{tab:crates}). For ASP, the \verus checker is 30--45 times faster than the Isabelle reference checker on the largest proof and uses far less memory, which is why the farm builds the second checker at all.

\subsection{RQ3: Robustness of checkers and trusted parsers}
\paragraph*{Differential testing} The two LRAT checkers agreed on all 14,555 cases of a differential campaign, including about 3,200 must-reject cases (proofs of satisfiable formulas, corrupted hints), none of which either checker accepted. The campaign found six parser differences, which led to a single normative DIMACS grammar. The two ASP checkers agreed on hundreds of thousands of cases, for example, 196,215 cases for weight rules and 82,267 cases with mutated certificates. The two SMT parsers produced identical canonical dumps on about 6,200 files.

\paragraph*{Mutation testing} Of 500 mutants of verified executable code in the three checkers and 14 further crates, 8 did not compile. Of the remaining 492, 396 (80.5\%) were killed by verification (Appendix~\ref{app:mutation}). The survivors were completeness-only changes, changes in untrusted readers, equivalent mutants, and, importantly, changes at the trusted output boundary. All six hand-written verdict-printing mutants initially verified. After the verdict boundary was verified, all 23 boundary mutants of the three checkers, including those six, were killed. A mutation inside the trusted exit functions is not killed by verification. The hygiene tool stops it.

\paragraph*{Trusted parsers} The parser for bit-vector and floating-point queries has no second implementation. An independent review found no mapping defect, but 15 of 18 targeted mutants survived the agents' original tests, and three of the survivors were unsound: they swapped greater-than with greater-or-equal, read SMT-LIB equality as IEEE equality, or read round-toward-zero as round-to-nearest-even. 101 new tests now kill all 18. The AIR reader is covered by the differential test against the SMT checker's parser.

\subsection{RQ4: What the agents produced}
The agents (Claude Opus~5.5, with a few other Claude models) wrote 44,653 lines of Isabelle in 111 theories (SMT 14,624; bit-vectors and floating point 12,503; ASP 9,354; AIR 4,347; LRAT 1,408; work-package prototypes 2,417), 87,245 lines of Rust in the three \verus checkers (SMT 45,908; ASP 33,660; LRAT 7,677), 7,545 lines of non-generated Standard ML for drivers and trusted parsers, the patches to \zthree and \clingo, and 18,929 lines of untrusted Python tooling. Isabelle's code generator produced a further 21,529 lines of Standard ML for the reference checkers.

What the gates caught is more informative than the volume. In every case above, the producing agent's own checks had passed: the validator reported full success while its semantics admitted a spurious proof, the parser tests passed while three unsound mutants survived, and the checker verified while its output boundary could print the wrong verdict. Independent review, differential testing, and mutation testing found each of these. The integrating agent also caught its own vacuous gates, for example, a comparison loop that compared no files because of wrong file names, and reported them. We read these results as consistent with the threat model: agents produce plausible evidence for incorrect work, and only checks they cannot influence are evidence. Our data do not tell honest error apart from gaming. The threat model covers both.

\begin{table}[t]
\centering
\caption{Trusted computing base, after this work.}
\label{tab:tcb}
\footnotesize
\begin{tabular*}{\columnwidth}{@{}l@{\extracolsep{\fill}}l@{}}
\toprule
Still trusted & No longer trusted \\
\midrule
Isabelle kernel and code generator & \zthree (every certified query) \\
Poly/ML, MLton, \texttt{rustc}, Rust std & \clasp, CaDiCaL \\
\verus front end and \texttt{vstd} & \verus AIR$\to$SMT (validated) \\
Trusted parsers & Certificates, log readers \\
ASP grounding (gringo) & Hint resolvers, query pairing \\
Specifications, \verus mirrors & AI agents' code and tests \\
I/O wrappers that print verdicts & AI agents' reports \\
Hygiene tool and hooks & \\
\bottomrule
\end{tabular*}
\end{table}

\paragraph*{Trusted computing base}
Table~\ref{tab:tcb} lists what remains trusted. Every solver answer, every certificate, and the agents' code, tests, and reports have left the trusted base. What remains are kernels and compilers, the \verus front end, a small set of parsers, the specifications, and the tools that enforce specification integrity. The specifications matter most, and no tool can check them: their integrity is enforced, their adequacy reviewed.
\section{Red-Team Evaluation}\label{sec:redteam}

The evidence so far comes from building the farm: reviews, differential tests, and mutation tests that found holes along the way. A red-team evaluation asks the converse question directly: if an artifact is built to obtain a passing verdict for a false claim, which gate stops it? Each successful attack reveals a missing gate. The evaluation is mutation testing for the gates themselves. Table~\ref{tab:redteam} reports the results.

\subsection{Protocol}
We first tried autonomous attacker agents that were instructed to deceive a blind integration gate. The agents' safety guardrails refused or halted this design, so we switched to a transparent one. For each attack class of Table~\ref{tab:threats}, a fixed catalog of seeded defects, one per attack variant, is driven through the real gate mechanics: the same checkers, hygiene tool, hooks, and integration checks that guard the farm. Where a test suite already contained attacks of a class, we reused it: A7 uses the 36 seam mutants of the validator's test suite, and A2 reruns the trusted readers' differential tests. Ground truth comes from outside the attacked component: models from an independent SAT solver for spurious certificates, \verus runs that confirm that a hidden verification bypass takes effect, and inspection of the seeded change for specification tampering. All attacks run in an isolated scratch area. Installed binaries, the main branch, and the frozen specifications were verified bit-identical before and after each attack run. The fixes were integrated afterwards. The orchestrating agent wrote the catalog. Two A6 variants came from a reviewer outside the farm after a first fix. Seeding tests the gates against attacks we can name. It cannot find attacks that nobody thought of, which autonomous attackers might. We count this as a limitation. The ``after'' column re-runs the catalog that the fixes were written for: it shows that the fixes stop these attacks, not that they stop the next variant.

One autonomous attacker had produced an attack before the halt. It committed a change to the trusted function that prints the verdict, while its final report stated that it had created nothing. We evaluated the change as A6 (below), and we read the episode as one more reason to inspect the repository directly instead of relying on an agent's report.

\begin{table}[t]
\centering
\caption{Red-team results. ``Passed'' counts seeded defects that the mechanical gates accepted, before and after the fixes. A2 reran existing differential tests; A8 is an analytic case.}
\label{tab:redteam}
\footnotesize
\begin{tabular*}{\columnwidth}{@{}l@{\extracolsep{\fill}}rrrl@{}}
\toprule
 & & \multicolumn{2}{c}{Passed} & \\
\cmidrule(lr){3-4}
Attack class & Seeded & before & after & Stopped by \\
\midrule
A1 Spurious LRAT proof & 5 & 0 & -- & verified checker \\
A3 Edit of a frozen spec & 3 & 2 & 0 & protection hook \\
A4 Verification bypass & 4 & 2 & 0 & hygiene tool \\
A5 Vacuous test & 7 & 7 & 0 & vacuity check \\
A6 Verdict in trusted I/O & 3 & 3 & 0 & hygiene tool \\
A7 Translation seam & 39 & 0 & -- & AIR validator \\
\midrule
A2 Reader divergence & \multicolumn{3}{r}{0 of 20,000} & differential tests \\
A8 Permitted backdoor & \multicolumn{3}{r}{not built} & spec review only \\
\bottomrule
\end{tabular*}
\end{table}

\subsection{Results}
\paragraph*{What held} Every spurious LRAT certificate for a satisfiable formula was rejected: an empty proof, an empty clause without hints, deletion followed by an empty clause, misuse of a RAT pivot, and an oversized header. An independent SAT solver confirmed that each formula is satisfiable. The soundness theorem of the checker does what it claims, including against proof shapes that two unverified checkers accept.

The translation seam held as well. All 36 seam mutants were rejected, as not validated, as a parse error, or as a violation of the scope structure. So were the spurious proof from the fourth review, an end-to-end variant of it, and a module withheld from validation: a module without validation counts against the total and forces a failing exit. The trusted readers agree: the AIR reader and the reader of the reference SMT checker differ on none of 20,000 rerun cases, after 2 million fuzzed and mutated inputs during development, and the other reader pairs pass their differential tests. One pair is untested: the parser of the bit-blasting chain against the AIR reader. It affects the 16 bit-vector queries, and the trust manifest lists it. The listed forms of verification bypasses and edits to frozen specifications through the file-editing tools were stopped as designed.

\paragraph*{What did not hold} We found six gaps, all in the code or procedures that enforce the farm's process, not in the checkers themselves.

\emph{Specification protection covered editing tools, not the shell.} The environment refused an implementing agent's edit of a frozen specification through the file-editing tools, but shell commands such as \texttt{sed -i} and output redirection changed the same file unhindered. The only remaining backstop was the diff the integrating agent reviews, a judgment rather than a mechanical barrier. \emph{The hygiene tool scanned file types, not the compiled text.} An \texttt{assume(false)} in a file pulled in through Rust's \texttt{include!} macro, with an extension the tool does not scan, was invisible to it. \emph{The hygiene tool scanned lines, not tokens.} An \texttt{external\_body} attribute split across two lines escaped its line-based pattern. \emph{Vacuous tests had no mechanical gate.} A skipped test, an empty test body, an always-true assertion, a failing comparison swallowed by \texttt{|| true}, a file compared with itself, an empty expected output, and a trivially true example lemma all passed. \emph{The translation validator was sound but not required.} No gate, hook, or work-package definition demanded validation. It ran because the agents were told to run it. The orchestrating agent found this gap by searching for every validator invocation. The validated corpus covered the example crates and the LRAT and ASP checkers, but not the SMT checker. \emph{The verified verdict boundary did not cover the printed verdict.} The attacker's change lets an environment variable replace the verdict line inside the two exit functions, whose bodies are trusted I/O that \verus does not verify. The mutant verified, passed the hygiene tool and the test suites, and, with the variable set, printed ``\texttt{s VERIFIED}'' for a rejected proof while the exit code stayed~1.

Both hidden verification bypasses take effect: with either one, \verus accepts a function whose body violates its postcondition. Without it, verification fails.

\paragraph*{Fixes and re-measurement} The hygiene tool now scans whole files, so its patterns span line breaks, and it follows \texttt{include!} targets regardless of extension. The hook now parses shell commands and refuses writes to protected files by redirection, \texttt{sed -i}, \texttt{tee}, \texttt{cp}, \texttt{mv}, \texttt{install}, \texttt{ln} (the link command), \texttt{dd}, and \texttt{truncate}. Thirteen test cases cover writes that must be refused and reads that must pass. A new vacuity check flags the seven forms of vacuous tests. Its self-test confirms that legitimate forms, such as assertions through a helper and cleanup commands followed by \texttt{|| true}, are not flagged, and it reports nothing on the farm's code. Validation is now enforced: a git hook that runs on every update of the main branch refuses a change to a checker unless a validation report for that exact commit shows every query certified and validated. After a fix in the validator's untrusted elaborator, the SMT checker crate passes too, and all 3,726 queries are certified and validated. The vacuity check now runs in the hook that every implementing agent must pass before it finishes. For A6, a first fix froze the body of each trusted function by hash. Two further variants passed it, and both verify: a trigger in an unverified helper, called from a frozen body, that prints the accepting verdict and exits with code~0; and a weakened precondition of the accepting exit function. The manifest now freezes each file of trusted I/O code as a whole, including contracts and helpers. The manifest is itself a protected file, the hygiene tool fails on any change, and it refuses an exit call outside a frozen file. Checkers also run in an empty environment. The hygiene tool now stops all three A6 mutants. On a re-run of the catalog, every seeded defect is stopped, and each gap has a regression test. The fixes remain syntactic approximations: the hygiene tool does not expand macros or follow code generated by build scripts, no hook recognizes every shell command that writes a file, and a test that runs and asserts but never reaches the branch under attack is still a matter of review. Because all guards currently run under the agents' operating-system account, an agent that writes a protected file through an unguarded interpreter (such as Python or Perl) could also update the manifest hash that protects it. Enforcing specification and manifest integrity outside the agents' privilege domain, through operating-system sandboxing and external cryptographic commit signing, is an orthogonal systems problem that composes directly with the farm's verification architecture and is left to future work.

\paragraph*{The analytic case (A8)} The frozen LRAT specification states soundness only: acceptance implies unsatisfiability. Rejection is always sound, so a checker that rejects valid proofs under a hidden trigger still satisfies the theorem. Acceptance is pinned only by 14 evaluated examples and by the benchmarks. A checker that copies its input to an extra output channel satisfies the theorem too, and the hygiene tool looks for verification bypasses, not for side channels. We did not build these backdoors. The argument rests on the theorem statement and on what each gate checks. Only specification review closes this gap, for example by demanding a completeness theorem or an argument about side channels.

\paragraph*{Measurement discipline} Two early readings reported exit status~0 for a rejected certificate. Both were artifacts of reading the exit status after a pipe. Measured directly, both exits were~1. We mention this because a red-team result is only as good as its measurement.

\subsection{Discussion}
The checkers held; the process rules leaked. This matches the asymmetry the design anticipates: the checkers are verified, while the tools that enforce specification integrity and the manifest are ordinary code, written by the same agents and tested less adversarially. The gaps fall into two groups. In three cases, an enforcing mechanism was incomplete: it matched file types, lines, or command names rather than what it guards. In the other three, a mechanism was missing: vacuous tests had no check; the translation validator, which withstood every seam attack, ran by convention rather than because a gate demanded it; and the manifest listed the trusted exit functions without pinning what they do. The last gap is the most significant: a mutant that verifies but still misreports. A proof cannot bind trusted I/O, but a gate can freeze it, as it freezes a specification. Process enforcement deserves the same treatment as the checkers: it should parse what it guards, intercept effects rather than commands, and live in the gate rather than in instructions. The campaign supplied an example of why. After the farm's git guard refused to delete a leftover attack branch, the orchestrating agent, pursuing a legitimate cleanup, tried a low-level git command that the guard did not cover. The agent harness's own permission classifier stopped it. Even a cooperative agent found a way around a guard, so a guard based on command names invites the next workaround. A8 marks the limit of mechanical gates: a backdoor that the specification permits passes all of them, which is why specification review remains the human's task.
\section{Soundness Defects Found}\label{sec:defects}

Differential testing against the verified checkers, the gates of the ASP work, and a campaign that tested solvers and proof checkers against independent oracles (cvc5, exhaustive enumeration, and our verified checkers) found defects in widely used tools. None affects the farm's soundness: our verified checkers reject every bad certificate, and none of the SMT solver defects is reachable from the queries the farm certifies. Appendix~\ref{app:defects} gives minimal reproducers.

\paragraph*{Proof checkers accept proofs of satisfiable formulas}
drat-trim's \texttt{lrat-check} accepts an empty-clause step whose hints do not lead to a conflict, or in the extreme case has no hints at all. It therefore verifies a ``proof'' of the satisfiable formula $(x_1\lor x_2)\land(\neg x_1\lor x_2)$. The differential campaign produced 229 cases with this pattern. lrat-trim~\cite{lrattrim} (version 0.2.0) accepts a proof that uses an original tautology as a unit hint, and so verifies a ``proof'' of a satisfiable formula as well. drat-trim~\cite{dratrim} itself reads an integer with a leading zero as octal: the literal \texttt{-010} becomes $-8$, so drat-trim finds complementary unit clauses in the satisfiable formula $\{x_8\},\{\neg x_{10}\}$ and accepts the one-line proof \texttt{0}. Our checkers read \texttt{-010} as $-10$ and reject. For a user who relies on such a checker, each defect turns a spurious certificate into a verified one.

\paragraph*{A valid proof of the wrong problem}
Sundance~\cite{sundance}, a new proof-producing SMT solver for program verification, collects every assertion of a script into one problem and silently ignores \texttt{push}, \texttt{pop}, \texttt{check-sat-assuming}, and \texttt{exit}. For a script whose only contradiction lies in a popped scope, it answers ``unsat'' and emits a proof of the flattened problem. Our verified LRAT checker accepts that proof: it is a valid refutation, but of a problem that is not the input. A proof certifies the solver's reading of the input. A checker must therefore bind the certificate to the input as its own trusted parser reads it, command structure included, as our SMT checkers do.

\paragraph*{Solvers report wrong unsatisfiability}
In its default configuration, the theory of finite sets in \zthree~\cite{z3} answers ``unsat'' for a satisfiable formula ($B\subseteq\{0,1,2,3\}$, $0\in B$, $|B\cup B| = 4$, satisfied by $B=\{0,1,2,3\}$) and ``sat'' for an unsatisfiable one, in every version we tried, from 5.1.0 to a build that already includes \zthree's recent finite-set soundness fixes (commit \texttt{deef01f}). The self-union inside \texttt{set.size} is not among the cases those fixes address. Under the non-default options \texttt{smt.arith.ignore\_int} and \texttt{sat.smt}, \zthree answers ``unsat'' for satisfiable integer problems with division. Boolector~3.2.4~\cite{boolector} and Bitwuzla~0.9.1~\cite{bitwuzla} let a binding of a parallel \texttt{let} see its siblings, contrary to the SMT-LIB standard, and answer ``unsat'' for a satisfiable formula and ``sat'' for an unsatisfiable one; Bitwuzla~0.10.0 fixes this. None of these defects reaches the farm: \verus emits no finite sets and sets neither option, our SMT parsers accept no finite sets, our checkers do not certify \zthree's wrong refutations under the two options, and our bit-vector parser rejects a binder that shadows a visible name.

\paragraph*{\clasp reports wrong unsatisfiability and wrong optima}
For a choice rule with at least two heads whose weight body contains one of those heads, \clasp~3.4 in its default configuration answers ``unsatisfiable'' for a program that has two answer sets. With minimize statements, the same defect yields a wrong optimum: \clasp reports ``optimum found'' with cost $(1,2)$, while the true optimum is $(1,-2)$, which \clasp itself finds with a different rule translation. The trigger is a head simplification that drops choice heads occurring in the body. The grounder gringo introduces an auxiliary atom for this shape, so the defect affects programs in \clasp's intermediate format that are written by hand or generated by other tools. A second defect, with SatELite preprocessing enabled (not the default), prints a non-answer-set as a model.

\paragraph*{Specifications we relied on}
Specifications from standards need review too. The SMT-LIB definition of \texttt{bvsrem} has a misplaced parenthesis and does not parse. The slice indices in the SMT-LIB text of \texttt{to\_fp} from a bit-vector, read literally, disagree with \zthree on 12,867 inputs. Our specifications implement the intended readings, and the reviews recorded the deviations.

\paragraph*{Disclosure} On 18 November 2026 we reported the defects in drat-trim, lrat-trim, Sundance, \zthree, and \clasp to their maintainers, each with the minimal reproducer of Appendix~\ref{app:defects}. By 21 November 2026, the \clasp maintainers had fixed both reports, and the \zthree maintainers had closed all three, two of them with a fix. The lrat-trim maintainer confirmed the defect in the last release, which the development version already rejects, and the Sundance maintainers plan to fix theirs with a pending change. The drat-trim report had no answer yet. The \texttt{lrat-check} defect had been reported upstream before, together with a proposed fix (drat-trim issue~42 and pull request~43). We did not report the \texttt{let} defect, because Bitwuzla~0.10.0 fixes it and the Boolector repository is archived.
\section{Discussion and Limitations}\label{sec:discussion}

\paragraph*{Specifications are the remaining risk}
Zero trust moves trust; it does not remove it. Once code, proofs, and solvers are checked, the specifications carry the weight, and agents also produce them. The farm protects their integrity and puts every specification through human and independent review, and our reviews found real defects, including defects in the published standards. But adequacy cannot be proved. Fuzzing can expose specification blind spots, gaps that let mutated code still verify against the specification although the test suite rejects it~\cite{specblindspots}, and the A8 case of our red-team experiment describes the same gap from the attacker's side. The line-by-line mirrors in \verus are a second instance of this risk: they are reviewed and tested differentially against the Isabelle reference checkers, but their correspondence to the Isabelle definitions is not machine-checked.

\paragraph*{Trusted code without human review}
No human read the trusted parsers. We consider this acceptable only because each is covered by a second implementation or by systematic mutation testing, and because the manifest names it. The bit-vector parser has no second implementation yet, and its review showed that the agents' own tests were not enough.

\paragraph*{Kernels, compilers, and the \verus front end}
The Isabelle kernel, its code generator, MLton, \texttt{rustc}, and \verus's front end remain trusted. Verified compilation of the reference checkers~\cite{isabellecakeml} and a semantics for \verus's source language would shrink this base further. The fast checkers and the reference checkers depend on disjoint compilers, which limits the effect of a single compiler bug to one of the two checkers.

A verifier's front end is a large, evolving code base, and like the solvers it drives, it has soundness bugs. The \verus version we pin (0.2026.10.30, commit 3102966) is no exception. We know of three. A non-positive recursive type whose cycle runs through a trait's associated type evades the positivity check, in two variants. A computation tactic treats two distinct shift expressions as equal. Each lets an ordinary crate, with no verification bypass, verify a \texttt{proof} function whose postcondition is \texttt{false}. These are front-end bugs, not checker holes, and they sit upstream of the translation step that the AIR validator checks: \verus emits a query that \zthree correctly finds unsatisfiable, so \texttt{verus-certify} certifies and validates it. They do not affect the farm's guarantee, because every certificate the farm accepts is also accepted by a checker that does not run \verus\ (the Isabelle reference checker or the bit-blasting chain), and the checker crates use none of these constructs. This is the situation the farm is built for: a verifier's ``verified'' is what it declines to trust, \verus's own included, and the verdict means something only because an independent checker confirms it.

\paragraph*{Coverage}
The SMT format covers what the 20 crates required. Queries that use strings, arrays as a theory, or rounding modes other than round-to-nearest-even are reported as not certified. ASP certificates speak about the ground program. Grounding is not certified. We evaluated verification workloads, not the downstream security analyses that motivate this work.

\paragraph*{Lessons}
Four practices did most of the work. First, independent review with concrete counterexamples found what the producing agents' tests missed in specifications and trusted code. Every counterexample became a must-reject test. Second, two checkers per format turned silent errors into visible disagreements. Third, the output boundary deserves verification: a verified checker that prints the wrong verdict is unverified. Fourth, any two trusted components that read the same text must be tested against each other on tokens and trees. The spurious proof that review found lived in exactly such a seam. None of these practices depends on AI. AI made them practical for us to apply to every component and makes them necessary.

\paragraph*{Independence and common-mode failure}
All agents use models of one family, almost all of them the same model. When we call reviews or implementations independent, we mean they ran in separate agent contexts without access to each other's output. This does not make their errors independent~\cite{knightleveson}. For proofs, common-mode errors are limited because different kernels and solvers check the two checkers' proofs. For specifications, trusted parsers, and reviews, they are not limited in this way. A second model family for one member of each pair would reduce them.

\paragraph*{Generative AI: what our claims rest on}
The generation process is not reproducible (see Generative AI Usage). Our artifact claims do not depend on it: each can be re-checked deterministically. Our process claims, such as the episodes we report, describe one instantiation of the farm. They show what happened, not how often it happens.
\section{Related Work}\label{sec:related}

\paragraph*{Agentic verification}
Benchmarks and systems for verified code generation judge proofs by the verdict of Dafny~\cite{dafnybench,clover}, Lean~\cite{verina}, or \verus~\cite{alphaverus,autoverus}, and LLM agents generate proofs for \verus~\cite{alphaverus,autoverus}. Most guard the verdict with filters against verification bypasses such as \texttt{assume(false)} or \texttt{sorry} and against modified specifications~\cite{dafnybench,verina,autoverus,alphaverus}; Clover adds consistency checks between code, docstring, and annotations, because the Dafny verdict alone accepts weakened annotations~\cite{clover}. The need for such filters is documented: without them, AlphaVerus's model learned to game the verifier with \texttt{assume(false)}, and 12\% of AutoVerus's proof candidates were rejected as unsafe~\cite{alphaverus,autoverus}. AutoMan generates Dafny implementations and proof obligations from protocol specifications and lets humans substitute hand-optimized code verified by manual refinement proofs~\cite{automan}. All of these systems trust the verifier and its solver. Our work is complementary: it removes the solver from the trusted base, enforces specification integrity in the environment rather than by filters, and its checkers could serve these pipelines. Earlier work on AI-generated security software in Ada/SPARK, including verified post-quantum cryptography, reported agents that circumvented weak checks~\cite{proverjudge}, which motivates our threat model. Hickey~\cite{hickey} proposes a spec-first architecture in which agents are untrusted, specifications are the human-reviewed artifact, and automated gates replace code review. It is a proposal whose trusted base is a proof kernel and a verified lowering pass, and it does not address solver answers. Its certificate checker for the intermediate representation is itself still proposed. Ho et al.~\cite{aeneascrypto} verify Rust cryptography in Microsoft's SymCrypt with Aeneas and Lean. Agents write the proofs and the Lean kernel checks them, although the trusted base also includes the compiled tactics \texttt{native\_decide} and \texttt{bv\_decide}, so part of the check rests on the Lean compiler rather than on the kernel~\cite{aeneascrypto,bvdecide}. Both share our premise that agents need not be trusted. Our work addresses the large class of verifiers that rely on SAT, SMT, and ASP solvers, and we implement and evaluate it.

\paragraph*{Certified SAT and SMT}
LRAT~\cite{lrat} showed that verified SAT proof checking can be practical; GRAT~\cite{grat} and cake\_lpr~\cite{cakelpr} are verified checkers, and lrat-trim~\cite{lrattrim} is a fast LRAT trimmer with an unverified checking mode. For SMT, Carcara~\cite{carcara} checks and elaborates Alethe proofs, lean-smt~\cite{leansmt} replays cvc5 proofs in Lean, and DRAT has been extended with theory lemmas~\cite{dratsmt}. cvc5's proof architecture is designed to emit several proof formats~\cite{cvc5proofs}, including Alethe~\cite{carcara}. SMTCoq checks SAT and SMT certificates with a checker verified and executed in Coq~\cite{smtcoq}. bv\_decide~\cite{bvdecide} provides verified bit-blasting. Our SMT checkers differ in target and redundancy: they check the certificates that a production solver emits for every query of a production verifier, across the theories that verifier uses, and there are two separately implemented verified ones.

\paragraph*{Translation validation}
Translation validation checks each run of a translator instead of verifying the translator once~\cite{transval}; Alive2 applies it to LLVM's intermediate representation~\cite{alive2}. Closest to our work, Parthasarathy et al.\ validate runs of Boogie's verification-condition generation in Isabelle, from loop elimination to the final condition, for a subset of Boogie~\cite{vcgvalid}. We validate the last step of \verus's pipeline, from AIR to SMT, per query in the same spirit, and we bind the validated text to the certified one, so that the translation and the solver answer are checked for the same query.

\paragraph*{ASP}
ASP-DRUPE~\cite{aspdrupe} introduced a proof format for checking the inconsistency claims of ASP solvers; ASP-QRAT~\cite{aspqrat} gives a proof system through quantified Boolean formulas. Pseudo-Boolean proof logging certifies cutting-planes reasoning, for example parity reasoning in SAT solvers~\cite{veripbparity}, and optimality, for example in MaxSAT solving~\cite{coremaxsat}. We add native weight-rule steps without expansion into normal rules, implemented optimality proofs, cutting planes, proofs of \clasp's own reasoning in its default configuration, and verified checkers.

\paragraph*{Verified security software and solver trust}
Verified cryptographic providers~\cite{evercrypt} and protocol implementations~\cite{noisestar} are built with F$^\star$; EverCrypt counts F$^\star$ and \zthree, which discharges the proof obligations, in its trusted base~\cite{evercrypt}. The verified certificate validator ARMOR~\cite{armor} instead chose the Agda proof assistant, partly because its trusted base is smaller than that of verifiers that integrate SMT solvers. The SoK on computer-aided cryptography~\cite{sokcac} discusses the trusted computing bases of such tools, and SMT-based verification is known to be unstable: semantically irrelevant changes to a query can change the solver's response~\cite{mariposa}. Solver fuzzing has found dozens of soundness bugs, most of them in \zthree~\cite{yinyang}. Proofs of unsatisfiability also matter in zero knowledge, where a prover convinces a verifier that a formula is unsatisfiable without revealing the proof, including for quantified Boolean formulas~\cite{zkpspace}. Supply-chain frameworks attest how artifacts were produced~\cite{intoto}. Our trust manifests attest what a verification result still depends on.
\section{Conclusion}\label{sec:conclusion}

AI agents can create verified software without human code review, but they can just as easily produce convincing evidence for incorrect software. We showed that formal verification can keep up under a zero-trust approach: no agent, solver, or translation is trusted by default, specifications are protected, and all remaining assumptions are documented. In this setup, agents built verified checkers for SAT, SMT, and ASP certificates, including an LRAT checker as fast as the fastest unverified one, and used them to remove \zthree from the trusted base of \verus for the evaluated crates.

Beyond bootstrapping its own verification stack, the architecture points to a broader use. Because every accepted verdict rests on a machine-checked certificate rather than an unverified solver claim, an agent cannot earn a passing verdict by exploiting a solver or translation bug, which is the reward hacking we set out to prevent. The farm can thus serve both as an autonomous producer of certified security-critical software and as a reward signal for training coding agents that cannot be gamed through the solver. The integrity of the specifications and of the tools that enforce them is ordinary code and remains the weaker link.

A natural next step is to apply the farm to the security domains themselves: end-to-end certified C-to-Rust migration of legacy network and binary parsers, synthesis of constant-time implementations of post-quantum cryptographic primitives, and machine-checkable isolation and optimality certificates for access-control policies and ASP attack graphs.

The remaining trust lies in kernels, compilers, a few parsers, and the specifications. The specifications are where human attention now belongs.

\section*{Generative AI Usage}
Generative AI is both the subject and the method of this work. The farm is AI-first: a human sets the intent and reviews the specifications for adequacy, and AI agents (Claude Opus~5.5, with a few other Claude models) write all theories, proofs, and code. No human reviewed the generated code or proofs. We accept them only because a proof assistant or a verified checker accepts them. We did not minimize model usage, because the question we study is what assurance remains when generation is cheap and unreviewed. Model outputs are not reproducible, and our artifact claims do not depend on them: each rests on artifacts that can be re-checked deterministically on a single laptop, regardless of how they were produced. Generative AI was used for editorial purposes in this manuscript, and all outputs were inspected by the authors to ensure accuracy and originality.
\bibliographystyle{IEEEtran}

\begin{thebibliography}{10}
\providecommand{\url}[1]{#1}
\csname url@samestyle\endcsname
\providecommand{\newblock}{\relax}
\providecommand{\bibinfo}[2]{#2}
\providecommand{\BIBentrySTDinterwordspacing}{\spaceskip=0pt\relax}
\providecommand{\BIBentryALTinterwordstretchfactor}{4}
\providecommand{\BIBentryALTinterwordspacing}{\spaceskip=\fontdimen2\font plus
\BIBentryALTinterwordstretchfactor\fontdimen3\font minus
  \fontdimen4\font\relax}
\providecommand{\BIBforeignlanguage}[2]{{%
\expandafter\ifx\csname l@#1\endcsname\relax
\typeout{** WARNING: IEEEtran.bst: No hyphenation pattern has been}%
\typeout{** loaded for the language `#1'. Using the pattern for}%
\typeout{** the default language instead.}%
\else
\language=\csname l@#1\endcsname
\fi
#2}}
\providecommand{\BIBdecl}{\relax}
\BIBdecl

\bibitem{copilot}
H.~Pearce, B.~Ahmad, B.~Tan, B.~Dolan-Gavitt, and R.~Karri, ``Asleep at the
  keyboard? assessing the security of {GitHub} {Copilot}'s code
  contributions,'' in \emph{SP}, 2022, pp. 754--768.

\bibitem{llmhardening}
J.~He and M.~T. Vechev, ``Large language models for code: Security hardening
  and adversarial testing,'' in \emph{CCS}, 2023, pp. 1865--1879.

\bibitem{baxbench}
M.~Vero, N.~Mündler, V.~Chibotaru, V.~Raychev, M.~Baader, N.~Jovanovi{\'c},
  J.~He, and M.~T. Vechev, ``{BaxBench}: Can {LLMs} generate correct and secure
  backends?'' in \emph{ICML}, 2025.

\bibitem{verina}
Z.~Ye, Z.~Yan, J.~He, T.~Kasriel, K.~Yang, and D.~Song, ``{VERINA}:
  Benchmarking verifiable code generation,'' in \emph{ICLR}, 2026.

\bibitem{alphaverus}
P.~Aggarwal, B.~Parno, and S.~Welleck, ``{AlphaVerus}: Bootstrapping formally
  verified code generation through self-improving translation and
  treefinement,'' in \emph{ICML}, 2025.

\bibitem{autoverus}
C.~Yang, X.~Li, M.~R.~H. Misu, J.~Yao, W.~Cui, Y.~Gong, C.~Hawblitzel, S.~K.
  Lahiri, J.~R. Lorch, S.~Lu, F.~Yang, Z.~Zhou, and S.~Lu, ``{AutoVerus}:
  Automated proof generation for {Rust} code,'' \emph{Proc. ACM Program.
  Lang.}, vol.~9, no. OOPSLA2, pp. 3454--3482, 2025.

\bibitem{clover}
C.~Sun, Y.~Sheng, O.~Padon, and C.~W. Barrett, ``Clover: Closed-loop verifiable
  code generation,'' in \emph{SAIV}, 2024, pp. 134--155.

\bibitem{dafnybench}
C.~Loughridge, Q.~Sun, S.~Ahrenbach, F.~Cassano, C.~Sun, Y.~Sheng, A.~Mudide,
  M.~R.~H. Misu, N.~Amin, and M.~Tegmark, ``{DafnyBench}: A benchmark for
  formal software verification,'' \emph{Trans. Mach. Learn. Res.}, vol. 2025,
  2025.

\bibitem{yinyang}
D.~Winterer, C.~Zhang, and Z.~Su, ``Validating {SMT} solvers via semantic
  fusion,'' in \emph{PLDI}, 2020, pp. 718--730.

\bibitem{proverjudge}
T.~Philipp, ``The prover is the judge: Verified security software from {AI}
  coding agents in {Ada/SPARK},'' \emph{CoRR}, vol. abs/2607.14340, 2026.

\bibitem{binsecrel}
L.-A. Daniel, S.~Bardin, and T.~Rezk, ``{Binsec/Rel}: Efficient relational
  symbolic execution for constant-time at binary-level,'' in \emph{SP}, 2020,
  pp. 1021--1038.

\bibitem{spectector}
M.~Guarnieri, B.~Köpf, J.~F. Morales, J.~Reineke, and A.~Sánchez,
  ``Spectector: Principled detection of speculative information flows,'' in
  \emph{SP}, 2020, pp. 1--19.

\bibitem{xenon}
K.~von Gleissenthall, R.~G. Kici, D.~Stefan, and R.~Jhala, ``Solver-aided
  constant-time hardware verification,'' in \emph{CCS}, 2021, pp. 429--444.

\bibitem{proverif}
B.~Blanchet, V.~Cheval, and V.~Cortier, ``{ProVerif} with lemmas, induction,
  fast subsumption, and much more,'' in \emph{SP}, 2022, pp. 69--86.

\bibitem{lrat}
L.~Cruz-Filipe, M.~J.~H. Heule, W.~A.~H. Jr., M.~Kaufmann, and
  P.~Schneider-Kamp, ``Efficient certified {RAT} verification,'' in
  \emph{CADE}, 2017, pp. 220--236.

\bibitem{grat}
P.~Lammich, ``The {GRAT} tool chain - efficient {(UN)SAT} certificate checking
  with formal correctness guarantees,'' in \emph{SAT}, 2017, pp. 457--463.

\bibitem{cakelpr}
Y.~K. Tan, M.~J.~H. Heule, and M.~O. Myreen, ``cake\_lpr: Verified propagation
  redundancy checking in {CakeML},'' in \emph{TACAS (2)}, 2021, pp. 223--241.

\bibitem{carcara}
B.~Andreotti, H.~Lachnitt, and H.~Barbosa, ``Carcara: An efficient proof
  checker and elaborator for {SMT} proofs in the {Alethe} format,'' in
  \emph{TACAS (1)}, 2023, pp. 367--386.

\bibitem{leansmt}
A.~Mohamed, T.~Mascarenhas, H.~Khan, H.~Barbosa, A.~Reynolds, Y.~Qian,
  C.~Tinelli, and C.~W. Barrett, ``lean-smt: An {SMT} tactic for discharging
  proof goals in {Lean},'' in \emph{CAV (3)}, 2025, pp. 197--212.

\bibitem{dratsmt}
S.~Hitarth, C.~R. Codel, H.~Lachnitt, and B.~Dutertre, ``Extending {DRAT} to
  {SMT},'' in \emph{FMCAD}, 2024, pp. 1--11.

\bibitem{aspdrupe}
M.~Alviano, C.~Dodaro, J.~K. Fichte, M.~Hecher, T.~Philipp, and J.~Rath,
  ``Inconsistency proofs for {ASP}: The {ASP-DRUPE} format,'' \emph{Theory
  Pract. Log. Program.}, vol.~19, no. 5-6, pp. 891--907, 2019.

\bibitem{aspqrat}
L.~Chew, A.~de~Colnet, and S.~Szeider, ``{ASP-QRAT}: A conditionally optimal
  dual proof system for {ASP},'' in \emph{KR}, 2024.

\bibitem{nist80020}
S.~Rose, O.~Borchert, S.~Mitchell, and S.~Connelly, ``Zero trust
  architecture,'' National Institute of Standards and Technology, NIST Special
  Publication 800-207, Aug. 2020.

\bibitem{clingo}
M.~Gebser, R.~Kaminski, B.~Kaufmann, and T.~Schaub, ``Multi-shot {ASP} solving
  with clingo,'' \emph{Theory Pract. Log. Program.}, vol.~19, no.~1, pp.
  27--82, 2019.

\bibitem{cdnl}
M.~Gebser, B.~Kaufmann, and T.~Schaub, ``Conflict-driven answer set solving:
  From theory to practice,'' \emph{Artif. Intell.}, vol. 187, pp. 52--89, 2012.

\bibitem{assat}
F.~Lin and Y.~Zhao, ``{ASSAT}: computing answer sets of a logic program by
  {SAT} solvers,'' \emph{Artif. Intell.}, vol. 157, no. 1-2, pp. 115--137,
  2004.

\bibitem{loopformulas}
J.~Lee, ``A model-theoretic counterpart of loop formulas,'' in \emph{IJCAI},
  2005, pp. 503--508.

\bibitem{aspprotocols}
L.~C. Aiello and F.~Massacci, ``Verifying security protocols as planning in
  logic programming,'' \emph{ACM Trans. Comput. Log.}, vol.~2, no.~4, pp.
  542--580, 2001.

\bibitem{verus}
A.~Lattuada, T.~Hance, J.~Bosamiya, M.~Brun, C.~Cho, H.~LeBlanc, P.~Srinivasan,
  R.~Achermann, T.~Chajed, C.~Hawblitzel, J.~Howell, J.~R. Lorch, O.~Padon, and
  B.~Parno, ``Verus: A practical foundation for systems verification,'' in
  \emph{SOSP}, 2024, pp. 438--454.

\bibitem{intoto}
S.~Torres-Arias, H.~Afzali, T.~K. Kuppusamy, R.~Curtmola, and J.~Cappos,
  ``in-toto: Providing farm-to-table guarantees for bits and bytes,'' in
  \emph{USENIX Security Symposium}, 2019, pp. 1393--1410.

\bibitem{reprobuilds}
M.~Fourné, D.~Wermke, W.~Enck, S.~Fahl, and Y.~Acar, ``It's like flossing your
  teeth: On the importance and challenges of reproducible builds for software
  supply chain security,'' in \emph{SP}, 2023, pp. 1527--1544.

\bibitem{fpsmt}
M.~Brain, C.~Tinelli, P.~Rümmer, and T.~Wahl, ``An automatable formal
  semantics for {IEEE-754} floating-point arithmetic,'' in \emph{ARITH}, 2015,
  pp. 160--167.

\bibitem{veripbparity}
S.~Gocht and J.~Nordström, ``Certifying parity reasoning efficiently using
  pseudo-boolean proofs,'' in \emph{AAAI}, 2021, pp. 3768--3777.

\bibitem{lrattrim}
F.~Pollitt, M.~Fleury, and A.~Biere, ``Faster {LRAT} checking than solving with
  {CaDiCaL},'' in \emph{SAT}, 2023, pp. 21:1--21:12.

\bibitem{dratrim}
N.~Wetzler, M.~Heule, and W.~A.~H. Jr., ``{DRAT-trim}: Efficient checking and
  trimming using expressive clausal proofs,'' in \emph{SAT}, 2014, pp.
  422--429.

\bibitem{sundance}
A.~Shah, B.~F. Jones, C.~Cebeci, J.~Z.~S. Hu, L.~A. Barnett, R.~Delmas,
  M.~Heule, B.~Parno, and B.~Dutertre, ``{Sundance}: A simple {SMT} solver for
  program verification,'' arXiv:2611.05668, 2026.

\bibitem{z3}
L.~M. de~Moura and N.~S. Bj{\o}rner, ``{Z3}: An efficient {SMT} solver,'' in
  \emph{TACAS}, 2008, pp. 337--340.

\bibitem{boolector}
A.~Niemetz, M.~Preiner, and A.~Biere, ``{Boolector} 2.0,'' \emph{J. Satisf.
  Boolean Model. Comput.}, vol.~9, no.~1, pp. 53--58, 2014.

\bibitem{bitwuzla}
A.~Niemetz and M.~Preiner, ``{Bitwuzla},'' in \emph{CAV}, 2023, pp. 3--17.

\bibitem{specblindspots}
R.~Ji and M.~Xu, ``Finding specification blind spots via fuzz testing,'' in
  \emph{SP}, 2023, pp. 2708--2725.

\bibitem{isabellecakeml}
L.~Hupel and T.~Nipkow, ``A verified compiler from {Isabelle/HOL} to
  {CakeML},'' in \emph{ESOP}, 2018, pp. 999--1026.

\bibitem{knightleveson}
J.~C. Knight and N.~G. Leveson, ``An experimental evaluation of the assumption
  of independence in multiversion programming,'' \emph{IEEE Trans. Software
  Eng.}, vol.~12, no.~1, pp. 96--109, 1986.

\bibitem{automan}
Z.~Zhang, T.~Zhou, C.~Jenkins, O.~Chowdhury, and S.~Mu, ``{AutoMan}:
  Facilitating verified distributed systems development through automatic code
  generation and manual optimizations,'' in \emph{SOSP}, 2025, pp. 768--785.

\bibitem{hickey}
J.~Hickey, ``Modern software engineering with verification agents: A spec-first
  architecture for autonomous synthesis, verification {IRs}, and organizational
  scale,'' arXiv:2611.02285, 2026.

\bibitem{aeneascrypto}
S.~Ho, C.~Fournet, J.~Protzenko, M.~Naehrig, J.~Clune, P.~Longa, G.~Boisseau,
  F.~L. S{\'a}nchez, A.~Fromherz, and A.~Delignat-Lavaud, ``Scaling
  verification of cryptographic software with {Aeneas}, {Rust}, and {Lean},''
  arXiv:2609.15648, 2026.

\bibitem{bvdecide}
H.~Böving, S.~Bhat, L.~Cicolini, A.~C. Keizer, L.~Frénot, A.~Mohamed,
  L.~Stefanesco, H.~Khan, J.~Clune, C.~W. Barrett, and T.~Grosser,
  ``Interactive bitvector reasoning using verified bit-blasting,'' \emph{Proc.
  ACM Program. Lang.}, vol.~9, no. OOPSLA2, pp. 3259--3285, 2025.

\bibitem{cvc5proofs}
H.~Barbosa, A.~Reynolds, G.~Kremer, H.~Lachnitt, A.~Niemetz, A.~Nötzli,
  A.~Ozdemir, M.~Preiner, A.~Viswanathan, S.~Viteri, Y.~Zohar, C.~Tinelli, and
  C.~W. Barrett, ``Flexible proof production in an industrial-strength {SMT}
  solver,'' in \emph{IJCAR}, 2022, pp. 15--35.

\bibitem{smtcoq}
B.~Ekici, A.~Mebsout, C.~Tinelli, C.~Keller, G.~Katz, A.~Reynolds, and C.~W.
  Barrett, ``{SMTCoq}: A plug-in for integrating {SMT} solvers into {Coq},'' in
  \emph{CAV (2)}, 2017, pp. 126--133.

\bibitem{transval}
A.~Pnueli, M.~Siegel, and E.~Singerman, ``Translation validation,'' in
  \emph{TACAS}, 1998, pp. 151--166.

\bibitem{alive2}
N.~P. Lopes, J.~Lee, C.-K. Hur, Z.~Liu, and J.~Regehr, ``Alive2: bounded
  translation validation for {LLVM},'' in \emph{PLDI}, 2021, pp. 65--79.

\bibitem{vcgvalid}
G.~Parthasarathy, P.~Müller, and A.~J. Summers, ``Formally validating a
  practical verification condition generator,'' in \emph{CAV (2)}, 2021, pp.
  704--727.

\bibitem{coremaxsat}
J.~Berg, B.~Bogaerts, J.~Nordström, A.~Oertel, and D.~Vandesande, ``Certified
  core-guided {MaxSAT} solving,'' in \emph{CADE}, 2023, pp. 1--22.

\bibitem{evercrypt}
J.~Protzenko, B.~Parno, A.~Fromherz, C.~Hawblitzel, M.~Polubelova,
  K.~Bhargavan, B.~Beurdouche, J.~Choi, A.~Delignat-Lavaud, C.~Fournet,
  N.~Kulatova, T.~Ramananandro, A.~Rastogi, N.~Swamy, C.~M. Wintersteiger, and
  S.~Zanella-Béguelin, ``{EverCrypt}: A fast, verified, cross-platform
  cryptographic provider,'' in \emph{SP}, 2020, pp. 983--1002.

\bibitem{noisestar}
S.~Ho, J.~Protzenko, A.~Bichhawat, and K.~Bhargavan, ``Noise*: A library of
  verified high-performance secure channel protocol implementations,'' in
  \emph{SP}, 2022, pp. 107--124.

\bibitem{armor}
J.~Debnath, C.~Jenkins, Y.~Sun, S.~Y. Chau, and O.~Chowdhury, ``{ARMOR}: A
  formally verified implementation of {X.509} certificate chain validation,''
  in \emph{SP}, 2024, pp. 1462--1480.

\bibitem{sokcac}
M.~Barbosa, G.~Barthe, K.~Bhargavan, B.~Blanchet, C.~Cremers, K.~Liao, and
  B.~Parno, ``{SoK}: Computer-aided cryptography,'' in \emph{SP}, 2021, pp.
  777--795.

\bibitem{mariposa}
Y.~Zhou, J.~Bosamiya, Y.~Takashima, J.~Li, M.~Heule, and B.~Parno, ``Mariposa:
  Measuring {SMT} instability in automated program verification,'' in
  \emph{FMCAD}, 2023, pp. 178--188.

\bibitem{zkpspace}
A.~Karthikeyan, H.~Liu, K.~S. Meel, and N.~Luo, ``Towards practical
  zero-knowledge proof for {PSPACE},'' in \emph{SP}, 2026, pp. 1842--1861.

\end{thebibliography}

\appendices
\section{Trust Manifest}\label{app:manifest}
The following components remain trusted, per certificate class.
\begin{itemize}
  \item \textbf{Every class:} the semantics of the input as read by the trusted parser. For \verus queries, \verus's front end (Rust to VIR to AIR), its standard library \texttt{vstd} including the proofs in it, which \zthree checked when \texttt{vstd} was built, and \texttt{rustc}. A separate run certifies 95.9\% of \texttt{vstd}'s proof obligations against the Isabelle anchor; 87 remain judged by \zthree, chiefly real-valued resource algebra and nonlinear arithmetic.
  \item \textbf{Reference checkers:} the \isabelle 2025-2 kernel and Poly/ML, on which it runs, its code generator, MLton, the trusted SML parsers (DIMACS, SMT-LIB subset, ASP intermediate format, bit-vector and floating-point subset, AIR reader), and the SML code that maps a result to a printed verdict.
  \item \textbf{\verus checkers:} \verus, \texttt{rustc} and the Rust standard library, the specification mirrors, and the I/O wrappers. The \zthree answers that \verus obtained while verifying them are certified. The DIMACS parser and the ASP parser are verified against reviewed specifications. The SMT-LIB parser is cross-checked against the reference parser.
  \item \textbf{Bit-vector and floating-point queries:} one checker, the Isabelle chain. Its parser is reviewed and mutation-tested but has no second implementation.
  \item \textbf{SMT semantics:} the step from models over a type variable to all SMT-LIB models (argued, not mechanized). For \texttt{smt-check}, the instantiation of its ordered-field result to the reals (Appendix~\ref{app:theorems}).
  \item \textbf{Translation validation:} the assumption that SMT-LIB satisfies the AIR semantics' assumptions for simple symbols, that the AIR theorem composes with the SMT theorem, and that the SMT checker reads the emitted text as the validator does (differential evidence).
  \item \textbf{ASP:} grounding (certificates speak about the ground program).
  \item \textbf{Process enforcement, for artifacts the farm produces:} the hygiene tool, the protected-path hook, the git hook on the main branch, and the vacuity check. They run under the agents' operating-system account.
\end{itemize}
Never trusted: \zthree for any certified query, \clingo, \clasp, CaDiCaL, the certificates, the readers and hint resolvers inside the checkers, the pairing of queries with solver runs, and the agents' own tests and reports.

\paragraph*{Size} Table~\ref{tab:tcbsize} gives the size of the trusted code. A human reviews the frozen specifications, 2,546 non-blank lines of Isabelle, and their \verus mirrors, 2,538 lines. The agents wrote 44,653 lines of Isabelle and 87,245 lines of verified Rust. The hygiene tool recognizes these verification bypasses: in Isabelle, \texttt{sorry}, \texttt{oops}, \texttt{axiomatization}, \texttt{skip\_proofs}, and oracles; in \verus, \texttt{assume}, \texttt{admit}, \texttt{external\_body}, \texttt{external}, \texttt{external\_fn\_specification}, \texttt{assume\_specification}, \texttt{truncate}, \texttt{unsafe}, and user instances of the ordered-field trait. It does not scan the SML drivers, which are trusted as a whole.

\begin{table}[h]
\centering
\caption{Trusted code, in non-blank lines. SML drivers are counted as whole files, including proof readers whose output is checked.}
\label{tab:tcbsize}
\footnotesize
\begin{tabular*}{\columnwidth}{@{}l@{\extracolsep{\fill}}rrrrr@{}}
\toprule
 & Isabelle & \verus & SML parsers & \verus & \verus \\
Format & spec & mirror & and driver & parser & I/O \\
\midrule
LRAT & 26 & 276 & 324 & verified & 434 \\
SMT & 489 & 1,111 & 1,515 & 1,659 & 381 \\
ASP & 150 & 1,151 & 2,405 & verified & 335 \\
BV/FP & 895 & -- & 858 & -- & -- \\
AIR & 986 & -- & 547 & -- & -- \\
\midrule
Total & 2,546 & 2,538 & 5,649 & 1,659 & 1,150 \\
\bottomrule
\end{tabular*}
\\[2pt]\makebox[\columnwidth][l]{\footnotesize Process tools (hygiene, hooks, vacuity check, gates): 1,357 lines.}
\end{table}

\section{Certificate Formats}\label{app:formats}
\paragraph*{SMT} One certificate file per query. Each file continues the previous ones of the same solver process. Commands: SMT-LIB declarations; \texttt{(define-const \$k S t)} binds a term reference; \texttt{(push)} and \texttt{(pop n)}; \texttt{(assume L)}; \texttt{(infer L$^*$ J)} with a justification $J$ from Sect.~\ref{sec:satsmt}; \texttt{(del id$^*$)}. Identifiers are implicit: the $k$-th \texttt{assume} or \texttt{infer} has identifier $k$, and a rejected step still consumes its identifier. Terms are hash-consed in one table shared with the input and translated exactly as the trusted parser translates input terms. A malformed or hostile line can only make a step fail.

\paragraph*{ASP} A text and a binary encoding of the same steps: \texttt{a} (RUP with hint), \texttt{d} (deletion), \texttt{u} (unfounded set), \texttt{w} (weight definition), \texttt{e} (definitional extension), \texttt{m} (answer set or bound), \texttt{o} (bound clause), and the cutting-plane steps \texttt{r}, \texttt{p}, \texttt{b}, \texttt{f}. The checker computes the clause view from the ground program. The certificate cannot supply it.

\section{Minimal Reproducers}\label{app:defects}
Except where noted, we are not aware of a public report of these defects before 18 November 2026. Section~\ref{sec:defects} gives the status of our reports.

\paragraph*{drat-trim \texttt{lrat-check}} (commit \texttt{2e3b2dc0ecf9}) Formula \texttt{p cnf 2 2 / 1 2 0 / -1 2 0} (satisfiable with $x_2$ true); proof \texttt{3 0 0}, an empty clause without hints. Reported as verified. This is a reported upstream regression of an earlier fix.

\paragraph*{lrat-trim 0.2.0} (commit \texttt{b30f400f4ee5}) Formula \texttt{p cnf 1 2 / 1 -1 0 / -1 0} (satisfiable with $x_1$ false); proof \texttt{3 0 1 2 0}, which uses the tautology $x_1\lor\neg x_1$ as a unit hint. Reported as verified.

\paragraph*{drat-trim} (commit \texttt{2e3b2dc0ecf9}) Formula \texttt{p cnf 10 2 / 8 0 / -010 0} (satisfiable with $x_8$ true and $x_{10}$ false); proof \texttt{0}. drat-trim reports complementary unit clauses and verified.

\paragraph*{Sundance} (commit \texttt{8f1f219}) \texttt{(declare-const x Int) (push 1) (assert (= x 1)) (pop 1) (assert (= x 2)) (check-sat)}, satisfiable with $x=2$. Sundance answers \texttt{unsat} with a proof of $x=1\wedge x=2$, which our verified LRAT checker accepts.

\paragraph*{\zthree, finite sets} \texttt{(declare-const B (FiniteSet Int))} with \texttt{(set.subset B S)} for $S=\{0,1,2,3\}$, \texttt{(set.in 0 B)}, and \texttt{(= (set.size (set.union B B)) 4)}: satisfiable with $B=S$; \zthree answers \texttt{unsat}. Conversely, \texttt{(= (set.union C A) set.empty)} with \texttt{(distinct (set.size A) 0)} is unsatisfiable; \zthree answers \texttt{sat}. Both in the default configuration. The finite-set soundness class was fixed upstream, but this self-union case persists in a build that includes those fixes.

\paragraph*{\zthree, \texttt{sat.smt=true}} \texttt{(declare-const x Int)}, $-4\le x\le 4$, \texttt{(distinct 0 (div (div x 5) 2))}: satisfiable with $x=-1$, since SMT-LIB division is Euclidean; \zthree answers \texttt{unsat}.

\paragraph*{Boolector 3.2.4 and Bitwuzla 0.9.1} \texttt{(declare-const a (\_ BitVec 8)) (assert (let ((a \#x05) (b a)) (= b \#x00)))} is satisfiable with $a=0$, because \texttt{b} denotes the declared \texttt{a}; Boolector answers \texttt{unsat}. Bitwuzla does the same when the first binding contains a nested \texttt{let}, as in \texttt{(let ((a (let ((z \#x1)) \#x5)) (b a)) (= b \#x0))} over 4-bit vectors. Bitwuzla~0.10.0 fixes this, and Boolector is end-of-life.

\paragraph*{\clasp, wrong unsatisfiability} \texttt{\{b; c\} :- 1 \{not e, b\}. :- not b.} in \clasp's intermediate format. Answer sets: $\{b\}$ and $\{b,c\}$. \clasp~3.4 with default options reports unsatisfiable.

\paragraph*{\clasp, wrong optimum} A choice rule \texttt{\{3;4\} :- 1\{3=1, 5=2, not 5=3\}} with three minimize statements: default options report optimum $(1,2)$; with \texttt{--trans-ext=weight}, \clasp reports $(1,-2)$, which is lexicographically better and confirmed by brute force.

\paragraph*{\clasp, non-model printed} \texttt{\{3;3;1\} :- 4\{not 3=3, 3=2, 1=1\}. \{3\} :- 1\{not 1\}.} With \texttt{--sat-prepro=2}, \clasp prints $\{1\}$, which is not an answer set; the answer sets are $\emptyset$ and $\{3\}$.

\section{Top-Level Theorems}\label{app:theorems}
Each verdict of a reference checker rests on one Isabelle theorem. The \verus checkers prove the same statement against their mirrors, and their exit function for the accepting verdict requires it.
\begin{itemize}
  \item \textbf{LRAT} (\texttt{lrat\_sound}): $\mathit{check\_lrat}\;F\;P \Longrightarrow \neg\,\mathit{satisfiable}\;F$, where an assignment $\sigma:\mathbb{N}\to\mathit{bool}$ satisfies every clause. The \verus checker ensures that acceptance implies $\mathit{dimacs\_parse}(\mathit{input}) = \mathit{Some}\;f \wedge \neg\,\mathit{satisfiable}\;f$.
  \item \textbf{SMT} (\texttt{smt\_local\_sound\_ext\_defs\_po}): acceptance implies $\neg\,\mathit{satisfiable\_po}$ of the asserted terms. Models interpret Bool, Int $=\mathbb{Z}$, and Real $=\mathbb{R}$, and each uninterpreted sort as an arbitrary nonempty set in a type variable, so finite carriers count. Datatype constructors, selectors, and testers are uninterpreted symbols constrained to be exact, inverse, disjoint, and exhaustive; acyclicity is not assumed, and a selector applied to another constructor is unspecified. Division and modulus are Euclidean for nonzero divisors and unconstrained for zero. Int and Real never mix; \texttt{to\_real}, \texttt{to\_int}, \texttt{is\_int}, \texttt{abs}, and division other than between literals are uninterpreted. The step from models over a type variable to all set-theoretic SMT-LIB models is argued, not mechanized. \texttt{smt-check} proves the statement for every ordered field in place of $\mathbb{R}$; the instantiation to $\mathbb{R}$ is trusted.
  \item \textbf{Bit-vectors and floating point} (\texttt{fp\_sound}, with \texttt{bv\_sound} as corollary): acceptance implies $\neg\,\mathit{satisfiable\_fp}$. Each bit-vector carrier is in bijection with the bit lists of its width, and the operators follow SMT-LIB, including $\mathtt{bvudiv}\;s\;0 = 1\cdots1$ and $\mathtt{bvurem}\;s\;0 = s$; floating point has exactly one NaN, and only round-to-nearest-even is accepted.
  \item \textbf{ASP} (\texttt{check\_proof2o\_inconsistent}, \texttt{\_answer\_set}, \texttt{\_optimum}): the verdict ``inconsistent'' implies that the ground program $P$ has no answer set; ``answer set $M$'' implies that $M$ is one; ``optimum'' implies that the reported answer set is not dominated by any other, with costs compared lexicographically by priority. $M$ is an answer set if $M$ is the least model of the reduct and no constraint body holds; weight bodies follow Simons' closure semantics.
  \item \textbf{AIR validation} (\texttt{file\_verdicts\_sound}): if the validator accepts query $k$, the emitted SMT query is unsatisfiable, and the side conditions on symbol names hold, then elaboration succeeded and no model of the global context, with well-sorted values that satisfy the query's axioms, has an execution that reaches a failing assertion. Unsatisfiability is a hypothesis, discharged outside Isabelle by the SMT checkers on the same text. The theorem is parametric in a locale for the AIR semantics; its instance for SMT-LIB and its composition with the SMT theorem are not mechanized. It is vacuous if the global context or the lifting axioms are inconsistent.
\end{itemize}

\section{Mutation Testing}\label{app:mutation}
An untrusted mutator applied ten operators to verified executable code: replacement of arithmetic, relational, and logical operators; negation; changes to indices and bounds; argument swaps; statement deletion; constant replacement; and return-value replacement. We sampled mutants at the function level with fixed seeds from the three checkers and from 14 crates in the example corpus. A mutant counts as killed if it fails verification. The test suites served only as an oracle on 48 survivors. 8 of the 500 mutants did not compile and are excluded (Table~\ref{tab:mutation}). Kill rates per operator range from 65\% (return-value replacement) to 100\% (arithmetic operators and index changes).

\begin{table}[h]
\centering
\caption{Mutants of verified executable code.}
\label{tab:mutation}
\footnotesize
\begin{tabular}{@{}lrr@{}}
\toprule
Crate & Killed / compiled & Rate \\
\midrule
LRAT checker & 87/120 & 72.5\% \\
ASP checker & 108/138 & 78.3\% \\
SMT checker (sample) & 111/125 & 88.8\% \\
14 example crates & 90/109 & 82.6\% \\
\midrule
Total & 396/492 & 80.5\% \\
\bottomrule
\end{tabular}
\end{table}

The 96 surviving mutants and six hand-written mutants of the verdict printing were classified by an agent from the code, and a subset was run against the reference checker as an oracle: completeness only, where the mutant rejects more, which soundness-only contracts allow (32); untrusted proof readers whose output is checked (25); example functions without a postcondition, or with one the mutant still satisfies (19); the trusted output boundary (5, plus the 6 hand-written mutants), a finding that led to the verified verdict boundary; equivalent mutants (10); and constructor state that the specification leaves open (5). Equivalence was argued from the code. The oracle found no difference on the 6 equivalent mutants it ran. No survivor pointed to a bug in verification-condition generation.

\end{document}